\documentclass[a4paper, parskip=full,11pts, bibliography=totocnumbered]{article}
\input{paper.sty}
\usepackage[T1]{fontenc} 
\usepackage{subcaption}
\newcommand{\rd}{R_{D}}
\newcommand{\rds}{R_{D^{*}}}
\newcommand{\rdsb}{R_{D^{(*)}}}

\begin{document}
\title{\textcolor{KBFIred}{Towards a viable scalar interpretation of $\rdsb$}\vspace{.3em}}

\author[]{Sean Fraser
\thanks{sean.fraser@kbfi.ee} }

\author[]{Carlo Marzo
\thanks{carlo.marzo@kbfi.ee}}

\author[]{Luca Marzola
\thanks{luca.marzola@cern.ch} }

\author[]{Martti Raidal
\thanks{martti.raidal@cern.ch} }

\author[]{Christian Spethmann
\thanks{christian.spethmann@kbfi.ee} }

\affil[]{\KBFI}

\vspace{1cm}
\date{{Dated: \today}\vspace{0.5cm}}

\maketitle
\begin{abstract}
	\noindent\rule[0.75ex]{\linewidth}{0.75pt}
\noindent
Recent measurements of semileptonic B-meson decays seemingly imply violations of lepton flavor universality beyond the Standard Model predictions. With three-level explanations based on extended Higgs sectors being strongly challenged by the measurements of the $B_c^-$ lifetime, new theories invoking leptoquark or vector fields appear as the only feasible answer. However, in this work we show that simple scalar extensions of the Standard Model still offer a possible solution to the $B$ physics puzzle, owing to sizeable loop-level corrections which mimic the effects of new vector contributions. Considering a simplified model characterised by a charged and a neutral scalar particle, we verify the compatibility of the observed $R_{D^{(*)}}$ signal with the relevant collider bounds. We also study an embedding of the simplified model into a three-Higgs-doublet framework, and  investigate its main phenomenological constraints. 

	\noindent\rule[0.5ex]{\linewidth}{0.75pt}
\end{abstract}

%===============================================================================
% BODY
%===============================================================================
%-------------------------------------------------------------------------------
\section{Introduction} % (fold)
\begin{figure}[t]
	\centering
    \includegraphics[width=.45\linewidth]{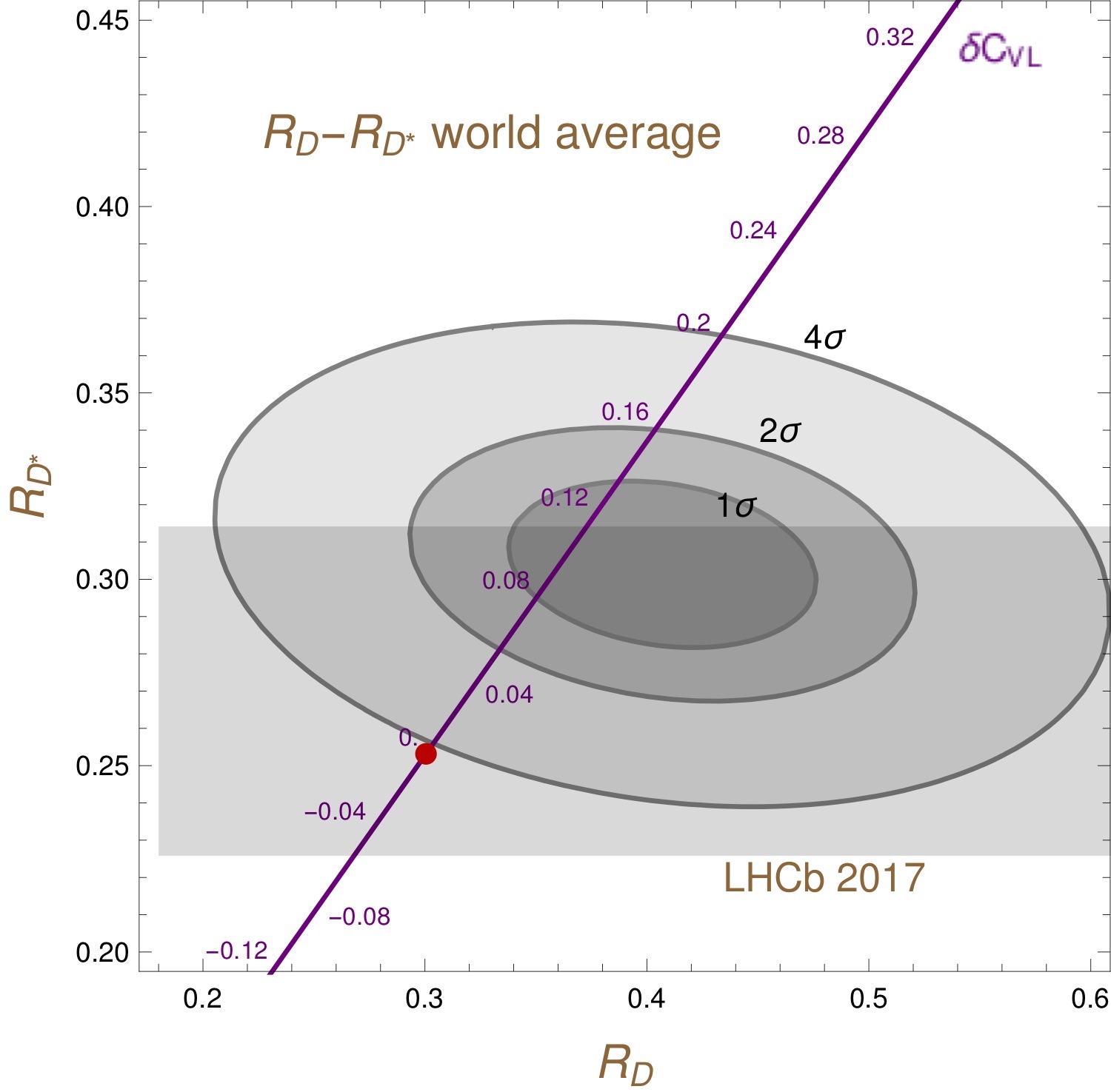}\,\,
	\caption{Hints of lepton flavour universality violation in the Heavy Flavour Average Group~\cite{HFAG} data. The red dot corresponds to the Standard Model prediction, whereas the horizontal grey band represents the region selected by the latest LHC$b$ results~\cite{Aaij:2017uff}. The purple solid line shows the dependence of $\rdsb$ on $\delta C_{VL} = -\delta C_{AL}$. $\delta C_{VL}$ is defined (similarly for $\delta C_{AL}$) as $\delta C_{VL} = C_{VL} - 1$ to parametrize the correction around the SM values $C_{VL}$=$-C_{AL}=1$.}
	\label{fig:operators}
\end{figure}

\label{sec:Introduction}
%-------------------------------------------------------------------------------
Recent results of the LHC$b$ experiment~\cite{Aaij:2015yra,Aaij:2017uff} highlight a significant amount of tension between the measured properties of the $B$-meson and their Standard Model (SM) predictions, confirming previous investigations by the BaBar~\cite{Lees:2012xj,Lees:2013uzd} and Belle~\cite{Huschle:2015rga,Sato:2016svk,Hirose:2016wfn} collaborations. 
In more detail, anomalies of lepton flavour universality have been measured for the underlying $b \to c \tau \bar{\nu}_\tau$ transition in the ratio of branching fractions
\begin{equation} 
R_{D^{(*)}} = \frac{\mathcal{B}(\bar{B} \to D^{(*)} \tau \bar{\nu})}{
\mathcal{B}(\bar{B} \to D^{(*)} \ell \bar{\nu})} \,,
\end{equation}
for $\ell=e,\mu$. Including the most recent LHC$b$ measurement, the Heavy Flavour Average Group (HFLAV) reports the world averages~\cite{HFAG}
\begin{eqnarray}
R_{D}^{\mathrm{exp}} &=& 
0.407 \pm 0.039 \pm 0.024 \,,
\nonumber \\
R_{D^*}^{\mathrm{exp}} &=& 
0.306 \pm 0.013 \pm 0.007 \,,
\end{eqnarray}
which exceed the SM predictions
\begin{eqnarray}
R_{D}^{\mathrm{SM}} &=& 
0.300 \pm 0.008 \,,
\nonumber \\
R_{D^*}^{\mathrm{SM}} &=& 
0.252 \pm 0.003 \,,
\end{eqnarray}
with a combined significance of about $4.1\sigma$. Being defined as ratios, $\rd$ and $\rds$ are quantities particularly suitable for investigating lepton universality regardless of hadronic uncertainties in the involved SM parameters. All independent probes of $B \to D^{(*)}$ transitions have exhibited an increased affinity for third generation leptons, resulting in a departure from universality that exceeds the SM predictions. The present situation is summarised in Fig.~\ref{fig:operators}, where the elliptical areas shaded in grey represent different joint confidence regions for the mentioned observables. The horizontal band indicates instead the range probed by the latest LHC$b$ result. The SM prediction, marked by a red dot, falls just on the outside of the 4$\sigma$ confidence region. 

New particles could clearly induce extra contributions to $\rdsb$, however most of the proposed interpretations are immediately falsified by correlated measurements in different experiments. For instance, scalar extensions of the SM that seek to simultaneously explain $\rdsb$ through a tree-level contribution inevitably leads to an enhancement of the decay $B_c^- \to \tau \bar\nu$, which is incompatible with the parameter space indicated by the $R_{D^*}$ anomaly~\cite{Crivellin:2012ye,Iguro:2017ysu,Lee:2017kbi,Alonso:2016oyd,Akeroyd:2017mhr,Martinez:2018ynq}. Once the strictest experimental bounds are considered, this incompatibility apparently rules out such interpretations of the signal for extended Higgs sector models, opening the way to alternative solutions involving leptoquark fields~\cite{Li:2016vvp,Calibbi:2017qbu}, extra-dimensions \cite{Megias:2016bde,Blanke:2018sro}, gauge extensions \cite{Asadi:2018wea,Greljo:2018ogz,Abdullah:2018ets} or specific supersymmetric models~\cite{Altmannshofer:2017poe}.

However, in this paper we demonstrate that scalar extensions of the SM are still viable and can relax, if not completely solve, the tensions due to the $B$-physics measurements. In essence, we propose the generation of effective operators typically associated with vector currents via scalar mediators at the loop level. Our results indicate that if such loop effects are considered on top of the SM contribution, the $R_{D^*}$ anomaly can be explained via a sizeable but perturbative correction.

The paper is organized as follows. In the upcoming section, we briefly review the operator basis used in the literature to discuss flavour anomalies. 
After that, in Sec.~\ref{sec:bounds} we introduce the simplified model and investigate its possible role in the measured flavour anomalies. In addition, we discuss the effects of the bounds cast by measurements of the $B_c$ and $\tau$ lifetimes, as well as the collider signatures of the proposed framework. In Sec.~\ref{sec:A possible UV completion} we identify a possible high-energy completion and review the main phenomenological constraints implied.
Finally, we summarise our findings in Sec.~\ref{sec:Conclusions}. 

%-------------------------------------------------------------------------------
\section{Flavour anomalies and Wilson coefficients} % (fold)
\label{sec:WC}
%-------------------------------------------------------------------------------
The low-energy behaviour investigated at LHC$b$ and in $B$-factories can be described by matching the framework at hand with an effective theory at the $W$ mass-scale and successively evolving the relevant quantities to the $b$-quark mass-scale via the renormalization group equations. The procedure results in a class of dimension six operators that preserve color and electric charge, with coefficients to be determined from the dynamics of the considered model \footnote{We simplify our work by neglecting Tensor operators, which play no role in our  analyses and by disregarding the running of the coefficients of vector operators, which receives a first non-vanishing QCD contribution only at the two-loop level~\cite{Gonzalez-Alonso:2017iyc,Bardhan:2016uhr}. }. Including right handed neutrinos, the effective Lagrangian for the process $b \rightarrow c \ell \bar{\nu}$ takes the form~\cite{Goldberger:1999yh,Cirigliano:2012ab,Dutta:2017wpq,Alok:2017qsi,Bardhan:2016uhr}

\begin{eqnarray} \label{effL}
\mathcal{L}_{eff}^{b \rightarrow c \ell \bar{\nu}} &=& \frac{2 G_F V_{cb}}{\sqrt{2}} \, \left( C^\ell_{VL}\,\mathcal{O}^\ell_{VL} + C^\ell_{AL}\,\mathcal{O}^\ell_{AL} + C^\ell_{SL}\,\mathcal{O}^\ell_{SL} + C^\ell_{PL}\,\mathcal{O}^\ell_{PL} + C^\ell_{VR}\,\mathcal{O}^\ell_{VR} \right. \nn \\  &+& \left. C^\ell_{AR}\,\mathcal{O}^\ell_{AR} + C^\ell_{SR}\,\mathcal{O}^\ell_{SR} + C^\ell_{PR}\,\mathcal{O}^\ell_{PR} \right) \,,
\end{eqnarray}
where the eight operators are given by 
\begin{eqnarray} \label{basis}
\mathcal{O}^\ell_{VX} &=& \left[\bar{c} \gamma^{\mu} b\right]\left[\bar{\ell}\gamma_{\mu} P_X \nu_{\ell}\right]\,, \\
\mathcal{O}^\ell_{AX} &=& \left[\bar{c} \gamma^{\mu} \gamma_5 b\right]\left[\bar{\ell}\gamma_{\mu} P_X \nu_{\ell}\right] \,, \\
\mathcal{O}^\ell_{SX} &=& \left[\bar{c} b\right]\left[\bar{\ell} P_X \nu_{\ell}\right]\,, \\
\mathcal{O}^\ell_{PX} &=& \left[\bar{c} \gamma_5 b\right]\left[\bar{\ell} P_X \nu_{\ell}\right] \,,
\end{eqnarray}

for $ X \in\{L,R\}$ and with $P_X$ being the corresponding chirality projector.

As mentioned before, the observables $\rdsb$ are defined as the ratios of the total branching fractions $\mathcal{B}_{\tau}^{D^{(*)}}(m_{\tau},\bold{C}^{\tau})$ for $B\to D^{(*)}\, \,\ell\,\nu_\ell$
\begin{eqnarray}
\rdsb = \frac{\mathcal{B}_{\tau}^{D^{(*)}}(m_{\tau},\bold{C}^{\tau})}{\mathcal{B}_{\mu/e}^{D^{(*)}}(m_{\mu/e},\bold{C}^{\mu/e})}\,,
\end{eqnarray} 
where the $B$ meson branching ratios are functions of the coefficients of the adopted operator basis: $\boldsymbol{C}^\ell = C^\ell_{VX}, C^\ell_{AX}, C^\ell_{SX}, C^\ell_{PX}$. 

In order to illustrate the reach of the approach, we schematise in Table~\ref{table1} the dependence of selected flavour observables on the Wilson coefficient of the effective Lagrangian. 
The reader is referred to Refs~\cite{Bardhan:2016uhr,Watanabe:2017mip,Bigi:2017jbd} for the relative hadronic form factors and for our analysis we will use the exact expressions given in ~\cite{Bardhan:2016uhr}.

\begin{table}[t]
	\centering 
\begin{tabular}{l|rrrrrrrrr}
	\toprule
	& $C_{VL}$ & $C_{AL}$ & $C_{SL}$ & $C_{PL}$ & $C_{VR}$ & $C_{AR}$ & $C_{SR}$ & $C_{PR}$\\
	\hline
	$R_{D}$ \rule{0ex}{2.5ex}     &     {\color{green}Yes}  &    {\color{red}No}   &   {\color{green}Yes}    &   {\color{red}No}  &   {\color{green}Yes}    &  {\color{red}No}   &   {\color{green}Yes}     &  {\color{red}No}   \\
	\hline
	$R_{D^*}$   \rule{0ex}{2.5ex} &  {\color{red}No}  &   {\color{green}Yes}  &   {\color{red}No}  &  {\color{green}Yes}   &    {\color{red}No}     &  {\color{green}Yes} &   {\color{red}No} &  {\color{green}Yes}     \\
	\hline
	%$R_{J/\psi}$  \rule{0ex}{2.5ex} &  {\color{green}Yes}  &   {\color{green}Yes}  &   {\color{red}No}  &  {\color{green}Yes}   &    {\color{red}No}     &  {\color{red}No} &   {\color{red}No} &  {\color{red}No}     \\
	%\hline
	$\Gamma_{B_c \rightarrow \tau \nu}$  \rule{0ex}{2.5ex}  &  {\color{red}No}  &   {\color{green}Yes}  &   {\color{red}No}  &  {\color{green}Yes}   &    {\color{red}No}     &  {\color{green}Yes} &   {\color{red}No} &  {\color{green}Yes}     \\
		\bottomrule
\end{tabular} 
	\caption{Schematic dependence of selected flavour observables on the Wilson coefficients of the effective theory at the $b$-quark mass-scale.}
	\label{table1}
\end{table}

The case of $\rdsb$ is shown in more detail in Fig.~\ref{fig:operators}. Clearly, non-vanishing values of the combination $C_{VL}$-$C_{AL}$ (indicated by a purple solid line) allow to ameliorate the agreement with observations. Notice that fitting the anomaly involves only modest deviations of the relevant Wilson coefficient from the corresponding SM values $C_{VL}$=$-C_{AL}=1$.

\begin{figure}[t!]
	\centering
	\begin{subfigure}[t]{0.45\textwidth}
		\centering
	\includegraphics[width=.8\linewidth]{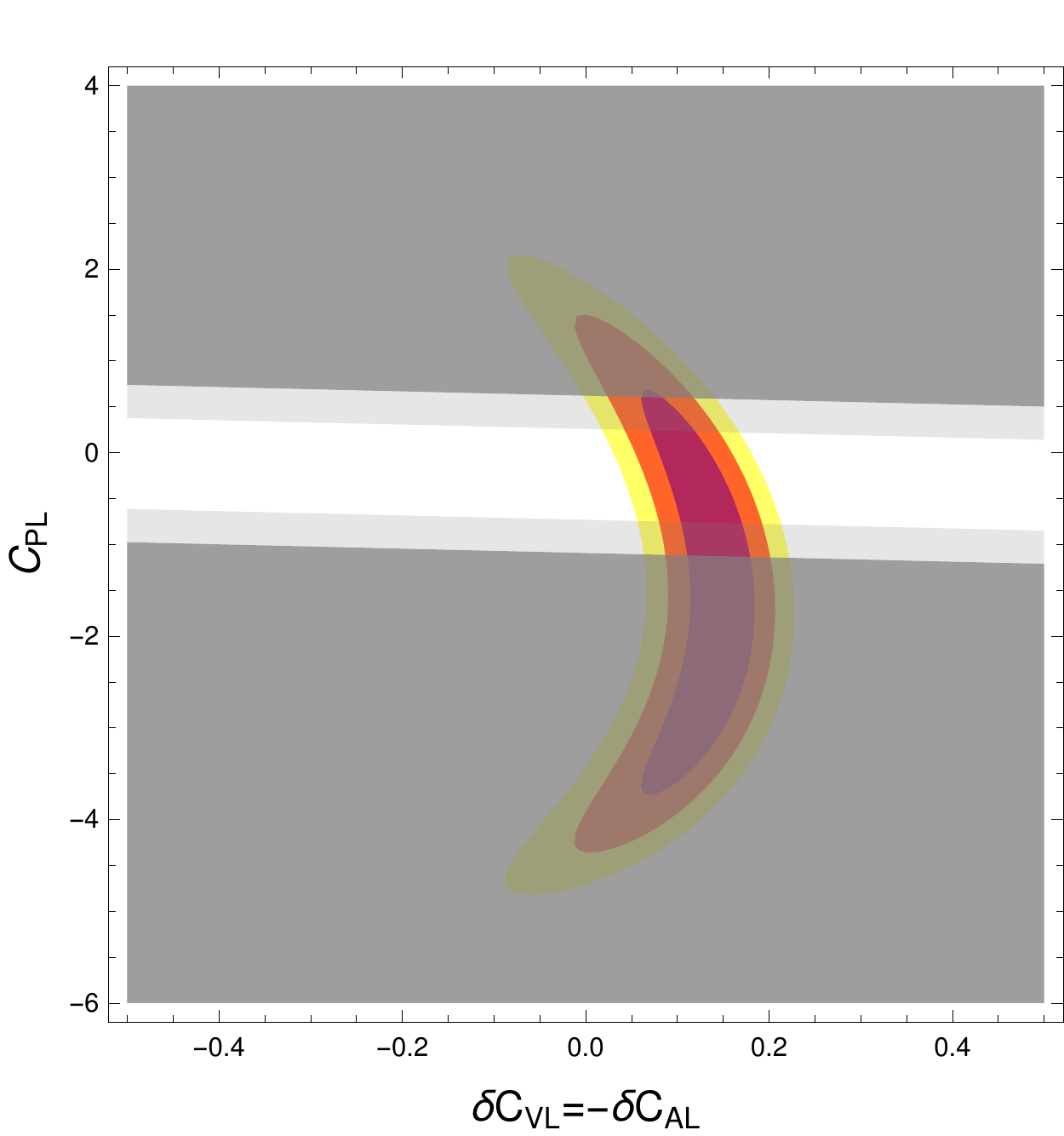}
	\caption{}
\label{fig:S2l}	
\end{subfigure}
\hspace{.2cm}
	\begin{subfigure}[t]{0.45\textwidth}
		\centering
	\includegraphics[width=.8\linewidth]{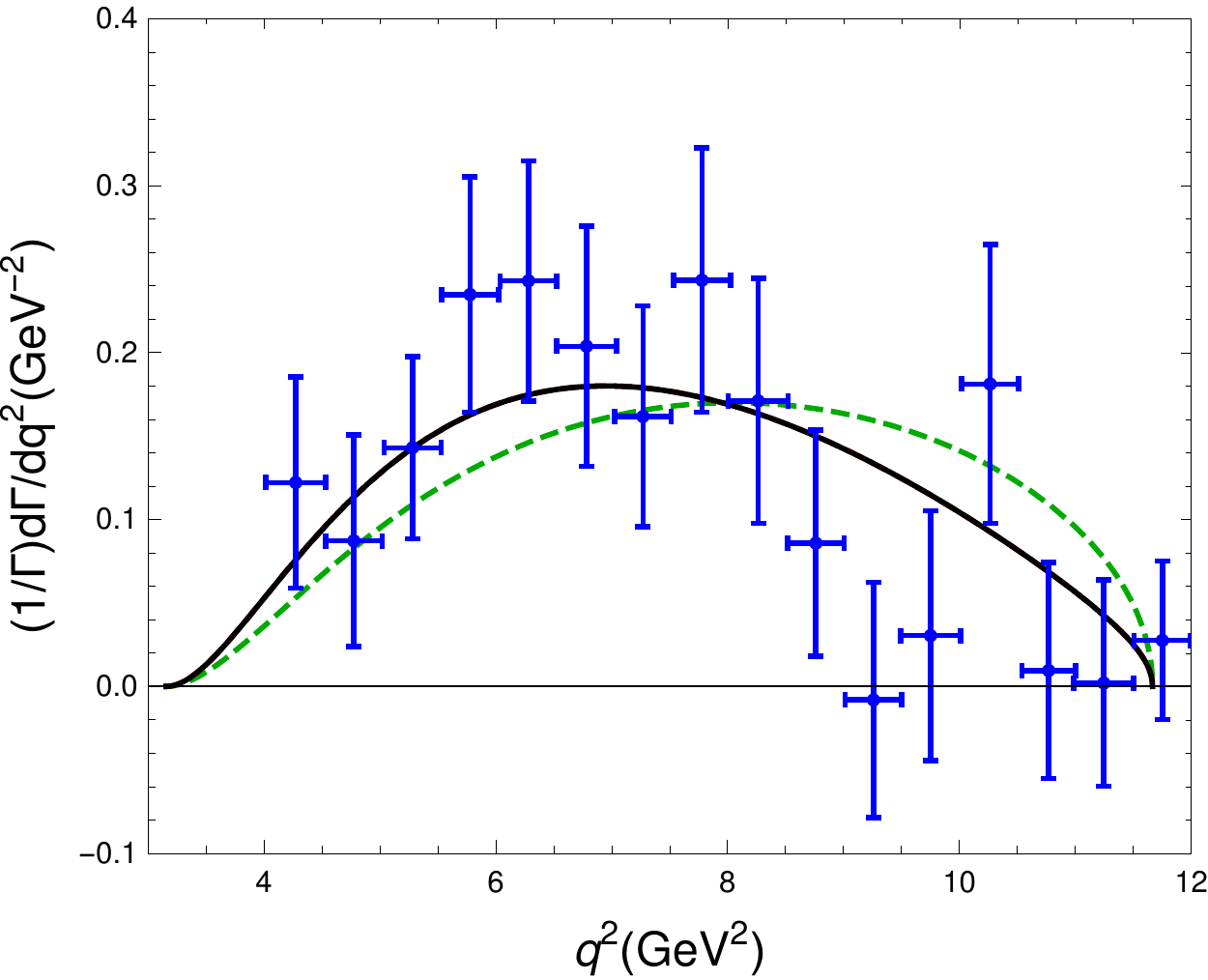}
	\caption{}
\label{fig:S2r}
	\end{subfigure}
	\caption{(a) The red, orange and yellow areas show the $1/2/3\sigma$  regions obtained for a joint fit of $R_D$ and $R_{D^*}$ as a function of the new physics contributions to $C_{VL} = - C_{AL}$ and $C_{PL}$. In correspondence to the light (dark) grey areas, the lifetime of $B_c$ deviates more than 10\% (30\%) form the measured value. (b) Differential branching ratio for the process $ B \to D\tau\nu$. The blue bars refer to the measurement gathered by the BABAR experiment~\cite{Lees:2013uzd}, whereas the black line represents the SM prediction, as well as that of models having only deviations in $C_{VL}$. The green dashed line is instead the prediction of a model characterised by $C_{SL}=0.35$, disfavoured by the experimental data.}
	\label{fig2}
\end{figure}

Traditional scalar extensions of the SM yield tree-level contributions to $C_{SL}$ and $ C_{PL}$, and, provided right handed neutrinos are considered, to their right-handed counterpart $C_{SR}$ and $ C_{PR}$. These coefficients are however severely constrained by measurements of the $B_c$ lifetime. In particular, the dependence of the branching ratio 
\begin{align}
\mathcal{B}_{\tau\nu}&=\tau_{B_c^-}\,\frac{m_{B_c}m_\tau^2 f_{B_c}^2G_F^2 |V_{cb}|^2}{8\pi}\left(1-\frac{m_\tau^2}{m_{B_c}^2}\right)^2\,\nn\\&
\left(\left|\frac{m_{B_c}^2}{m_\tau(m_b+m_c)}C_{PL}- C_{AL}\right|^2 + \left|\frac{m_{B_c}^2}{m_\tau(m_b+m_c)}C_{PR} - C_{AR}\right|^2\right)\,,
\label{Bctaunu} 
\end{align}   
for $B_c^-\to\tau\bar\nu_\tau$ on $C_{PL}$ is enhanced by the $m_{B_c}/m_\tau$ mass ratio. The impact of the $B_c$ lifetime measurements is illustrated in more detail in Fig.~\ref{fig:S2l}, where the red, orange and yellow areas show the $1/2/3\sigma$  regions for the joint fit of $R_D$ and $R_{D^*}$. The shaded areas indicate the values of the Wilson coefficients causing deviations larger than $10\%$ (light gray) or $30\%$ (dark gray) from the measured $B_c$ lifetime. As we can see, the latter is fully consistent with the SM prediction, $C_{SL} = C_{PL}$ = 0, and precludes explanations of the $\rdsb$ anomaly based on the $\mathcal{O}_{PL}$ operator. 

Measurements of the $B \to D\tau\nu$ decay spectrum provide instead a complementary way to probe the simplest scalar extensions~\cite{Lees:2013uzd}. As shown in Fig.~\ref{fig:S2r}, this observable constrains the scalar Wilson coefficient $C_{SL}$ but is essentially independent of the remaining operators in the effective Lagrangian~\eqref{effL}. Hence, models that for instance induce deviations only in the coefficient $C_{VL}$ result in predictions for $ B \to D\tau\nu$ indistinguishable from the SM one.    

Given the importance of the constraints cast by the measurements of the $B_c$ lifetime and by the $q^2$ distribution of $B \to D\tau\nu$, in the following we propose an effective scalar extension of the SM that explains the $\rdsb$ anomalies by generating sizeable contributions to $C_{VL}$ and keeps the coefficients $C_{SL}$ and $C_{PL}$ at their SM values. In our analyses we  refer to the 10\% limit on the $B_c$ lifetime when assessing the viability of the proposed scheme. To conclude the section, we mention that angular~\cite{Alok:2018uft} and differential distributions~\cite{Freytsis:2015qca} have shown further concrete possibilities to rule out some of the solutions of the $R_{D^{(*)}}$ puzzle.

%-------------------------------------------------------------------------------
\section{A simplified scalar model for flavour anomalies} % (fold)
\label{sec:bounds}
%-------------------------------------------------------------------------------

\begin{figure}[t]
	\centering
	\includegraphics[scale=0.8]{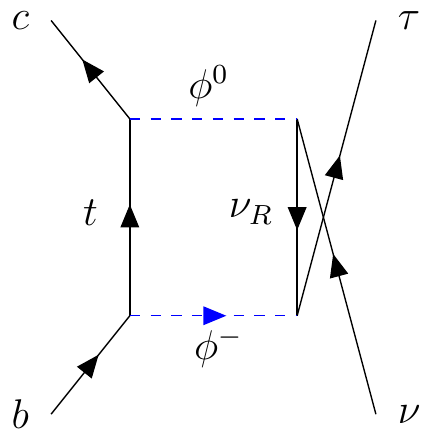}\hspace{.5cm}
	\includegraphics[scale=0.8]{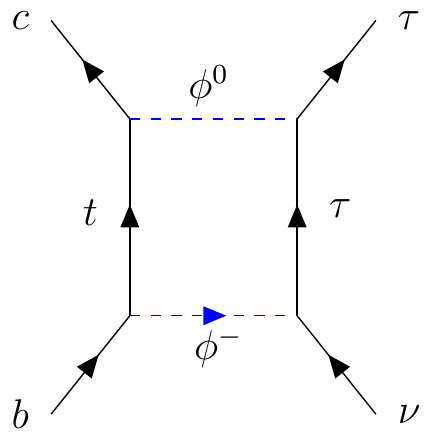}
	\caption{Additional one-loop diagrams for the $B_c$ meson decay induced by the simplified scalar model in Eq.~\eqref{LagNP}. The corresponding contributions are respectively quantified in Eq.~\eqref{cc1} and Eq.~\eqref{cc2}.  }
	\label{fig:S4}
\end{figure}

As mentioned in the previous section, scalar extensions of the SM with sizeable tree-level contributions cannot account for the current flavour anomalies. In order to investigate the effect of one-loop processes in the B-meson decay, we then focus on the following simplified model

\begin{eqnarray}
- \mathcal{L} &\supset& \xi_{\bar{c}_L t_R} \phi^{0*} \bar{c}_L t_R + \xi_{\bar{t}_R b_L} \phi^+ \bar{t}_R b_L + \xi_{\bar{\tau}_L \tau_R} \phi^{0} \bar{\tau}_L \tau_R - \xi_{\bar{\tau}_R \nu_L} \phi^-\bar{\tau}_R\,\nu_L + \nn \\ &+&  \xi_{\bar{\tau}_L\nu_R}\phi^- \bar{\tau}_L\nu_R  + \xi_{\bar{\nu}_R\nu_L} \phi^{0}\bar{\nu}_R\nu_L + h.c.  \, ,
\label{LagNP}
\end{eqnarray}

which introduces both an electrically neutral and a charged complex scalar field. The interactions contained in Eq.~\eqref{LagNP} have a non-zero projection on the vector and axial vector operators $O_{VL}$ and $O_{AL}$, providing new contributions to the relative Wilson coefficients via the diagrams shown in Fig.~\ref{fig:S4}: 

\begin{eqnarray}
\label{cc1}
&&C^1_{VL} = -C^1_{AL} = -\left(\frac{m_W^2}{8\pi^2 V_{cb} g_w^2}\right)\left(\xi_{c_L t_R} \xi_{t_R b_L} \xi_{\nu_R \nu_L} \xi_{\tau_L \nu_R} \right) D_{dd00}[m_{\nu_R}^2, m_t^2,m_{\phi^0}^2,m_{\phi^-}^2] 
\\
\label{cc2}
&&C^2_{VL} = -C^2_{AL} =  -\left(\frac{m_W^2}{8\pi^2 V_{cb} g_w^2}\right) \left(\xi_{c_L t_R} \xi_{t_R b_L} \xi_{\tau_L \tau_R} \xi_{\tau_R \nu_L} \right) D_{dd00}[m_t^2,m_{\phi^0}^2,m_{\phi^-}^2, m_\tau^2]\,.
\end{eqnarray}

In Eq.~\eqref{cc1} and in the rest of this work we follow the conventions of Ref.~\cite{Denner:1991kt} for the Passarino-Veltman integrals, as provided by Ref.s~\cite{Hahn:2000kx,Hahn:1998yk}. The given integrals are therefore the coefficients of the Lorentz-covariant tensor integrals. This means, in particular, that the relevant scalar integral of Eq.~\eqref{cc1} has to be identified with 

\begin{eqnarray}
\label{PaVe}
&& D_{dd00}[m_{1}^2, m_2^2,m_{3}^2,m_{4}^2] = \frac{\left(2\pi\mu \right)^{4-D}}{4\,i\pi^2}\int d^D q \frac{q^2}{\left(q^2-m_1^2\right)\left(q^2-m_2^2\right)\left(q^2-m_3^2\right)\left(q^2-m_4^2\right)}  \,.
\end{eqnarray}

The model includes also a sterile right-handed neutrino $\nu_R$ that plays an active role in the decay of the $B_c$ meson. We assume for our calculation   that the masses of these particles are negligible but still large enough to evade the cosmological bounds on additional relativistic species. We remark that the results we present are independent of the details of the neutrino mass generation mechanism as long as the interactions of $\nu_R$ are not significantly diminished by mixing effects.

Given the structure of the amplitudes for the diagrams in Fig.~\ref{fig:S4} and the overall normalization of the SM contribution proportional to $V_{cb} \sim 0.04$, we can produce percent-level variations in $C_{VL} = - C_{AL}$ for new scalar fields with masses in the range $100-500$ GeV and perturbative values of the couplings. Larger values would require the loss of perturbativity when fitting the  $R_{D^{(*)}}$ anomaly.
 
In order to assess the viability of the simplified model at hand, we investigate its collider phenomenology in conjunction with the signal produced from $B$ meson decays. After that, we discuss the details of a possible high-energy completion of the framework and the potential additional constraints it entails.

\subsection{Collider phenomenology of the simplified model}

The proposed model contains neutral and charged scalars, which can be singly produced at the LHC in association with top quarks~\cite{Akeroyd:2016ymd}. The initial state for such topologies contains $c$ and $b$ quarks, which are suppressed by their small contributions to the proton parton distribution functions at large momentum fractions. Because of this, the production cross section for single scalars and top quarks falls off steeply with increasing scalar mass. At the same time, however, the events become more visible as the transverse momentum of the decay products, as well as the missing energy from neutrinos, increases. In order to derive experimental limits for our model, we therefore scan the relevant parameter space to estimate the excluded region. 

In addition to single scalar/mono-top events, the charged scalars can be pair produced from quark/antiquark initial states through s-channel photon and t-channel top quark diagrams. The neutral scalars can also be pair produced via the same t-channel topology, resulting in particular in $4\tau$ and $4\nu$ final states, the latter being identifiable in mono-jet searches. Because the final states involve tau leptons, the suppressed electroweak processes are difficult to identify in the strongly interacting hadronic environment of the LHC. We therefore included the pair production events in the general scan over the model parameters.

\begin{figure}[t]
  \centering
    \includegraphics[width=.3\textwidth]{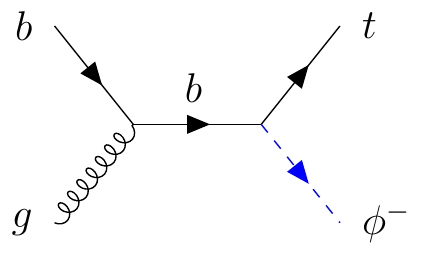}
	\includegraphics[width=.3\textwidth]{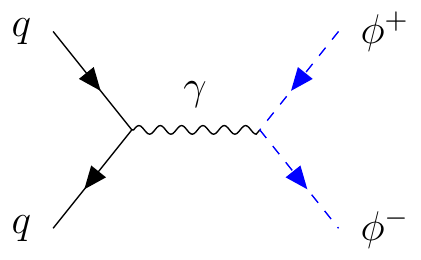}
	\includegraphics[width=.3\textwidth]{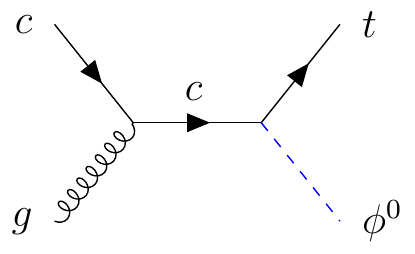}
	\\
	\includegraphics[width=.2\textwidth]{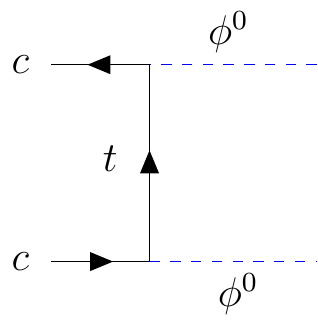}
	\hspace{.5cm}
	\includegraphics[width=.2\textwidth]{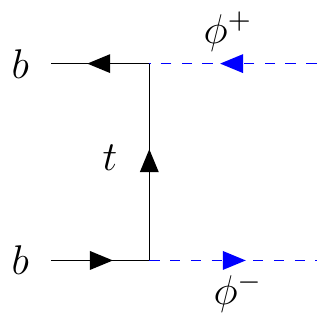}
  \caption{The considered scalar production topologies at the LHC.}
  \label{fig:tope}
\end{figure}

The topologies considered for the production of the charged and neutral scalars are shown in Fig.~\ref{fig:tope}. 
We decided to disregard the hadronic decay channels of the scalar bosons since events with decays to top quarks are kinematically suppressed, and decays to charm quarks cannot be identified over the QCD background. The analysed final states are reported in Table~\ref{tab:fs}.

\begin{table}[h!]
	\centering
	\begin{tabularx}{1\textwidth}{X<\centering | X<\centering X<\centering}
		  \toprule
		  & Charged scalar field & Neutral scalar field \\
		\hline
		Single production \rule{0ex}{2.5ex} &$t\,\tau\, \bar\nu $, $\bar t\,\bar\tau\, \nu $ &  $t\,\nu\,\bar\nu $, $t\,\tau\,\bar\tau$ \\
		Double production \rule{0ex}{2.5ex} & $\nu\, \bar\nu\,\tau\,\bar\tau$ & $\nu\,\bar\nu\,\nu\,\bar\nu\,j$, $\tau\,\bar\tau\,\tau\,\bar\tau$\\
		\bottomrule
	\end{tabularx}
	\caption{The analysed final states.}
	\label{tab:fs} 
\end{table}

In order to derive the bounds presented below, we implemented the Lagrangian for the proposed simplified model with \texttt{FeynRules}~\cite{Alloul:2013bka} and exported the resulting model files to \texttt{MadGraph5\_aMC@NLO}~\cite{Alwall:2014hca} via the \texttt{UFO} interface. For each of the topologies, we then performed a grid scan of the parameter space via Monte-Carlo simulations, applying the default kinematic cuts and collecting $10^5$ events per point. Finally, the partonic LHE files generated were compared against the 8 TeV analysis available in \texttt{CheckMATE 2.0.14}~\cite{Dercks:2016npn}. In addition, we seperately included the constraints presented in the dedicated CMS search~\cite{Khachatryan2015} for the single production of a charged Higgs boson.

Based on our scan of the parameter space, we now present three benchmark points.

\begin{figure}[t]
	\begin{eqnarray}
	&&\includegraphics[scale=0.51]{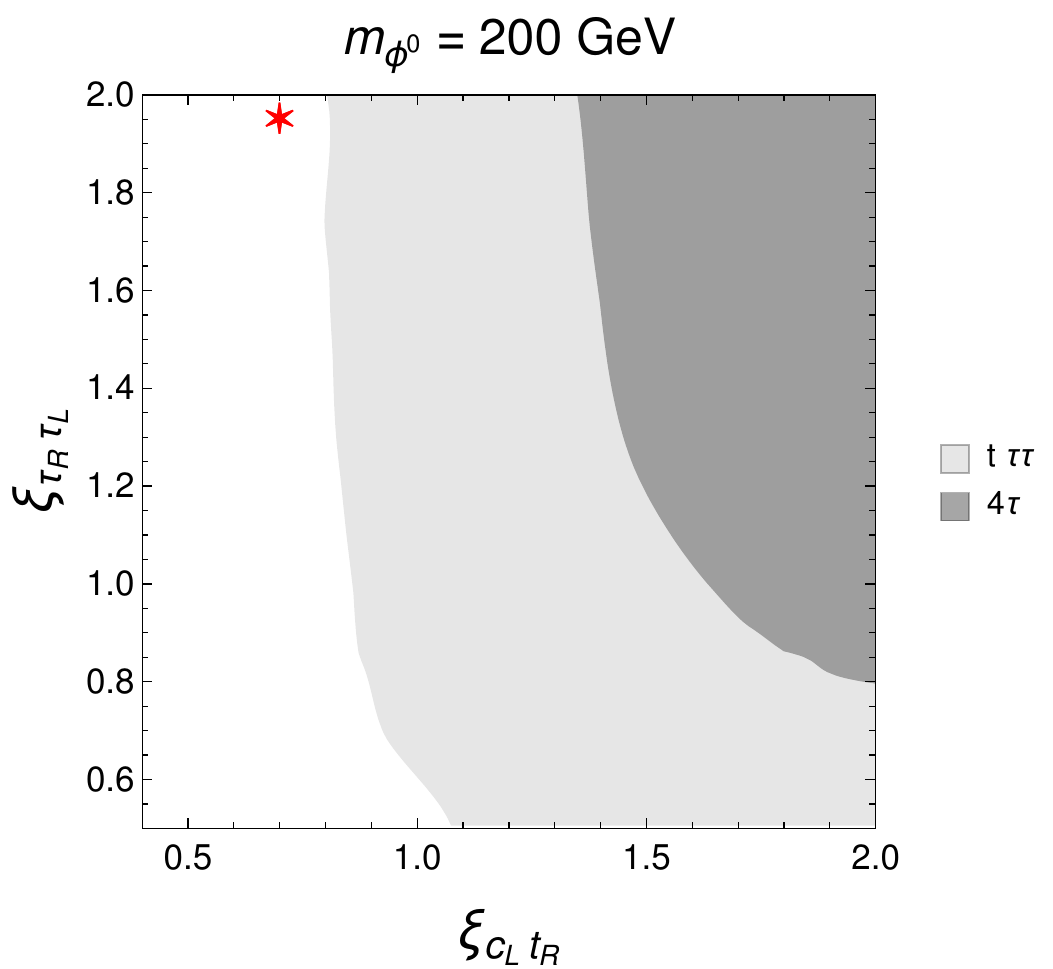} \hspace{2.2cm}
\includegraphics[scale=0.51]{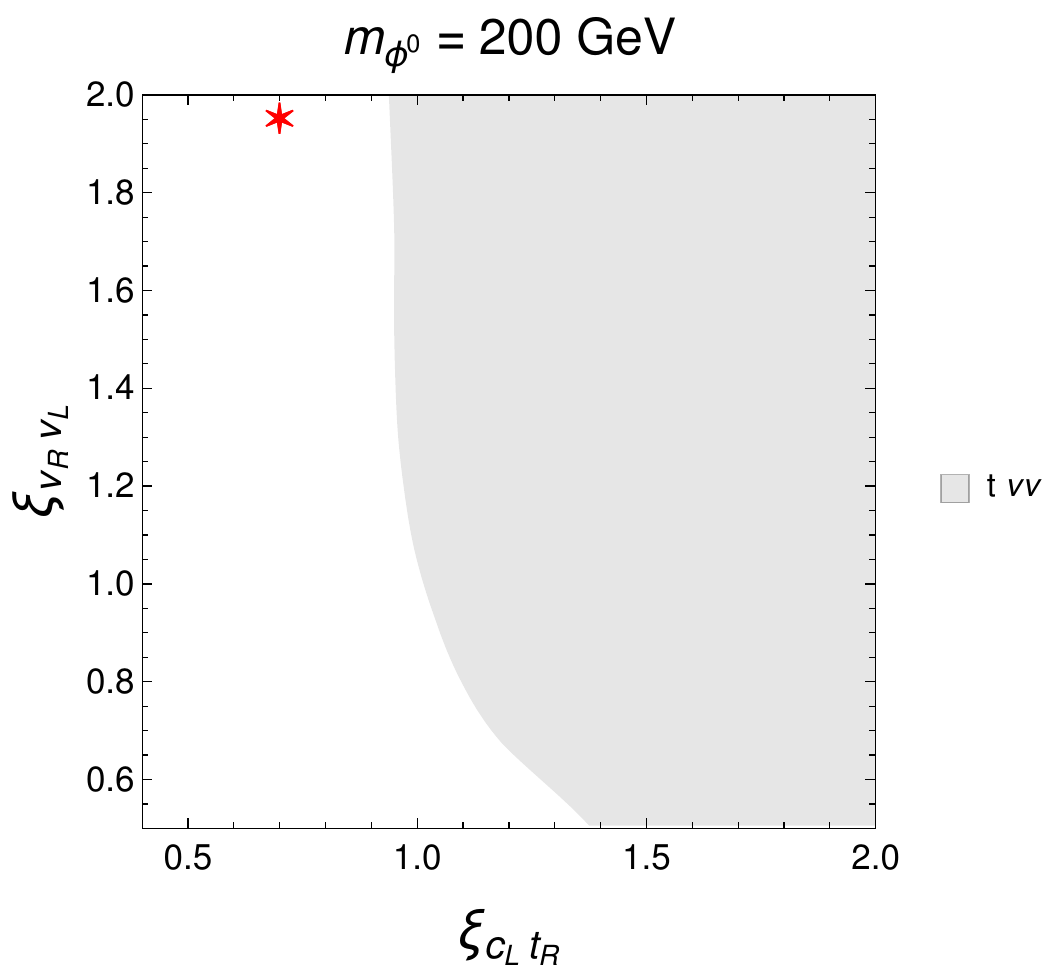}
\nn \\	&&\includegraphics[scale=0.5]{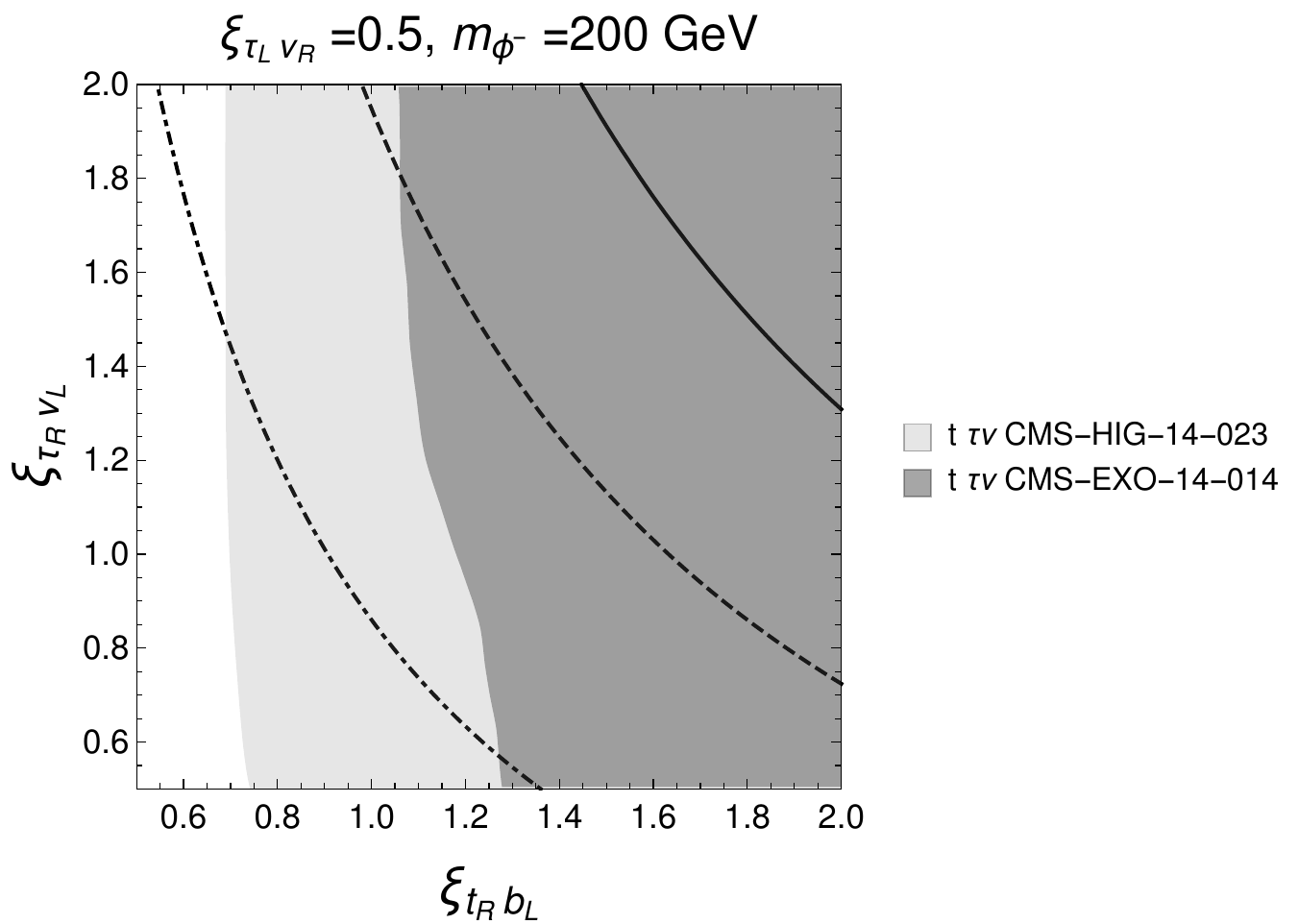}\,\,\,\,\,\,\,\,\,\,\,\,\,
\includegraphics[scale=0.5]{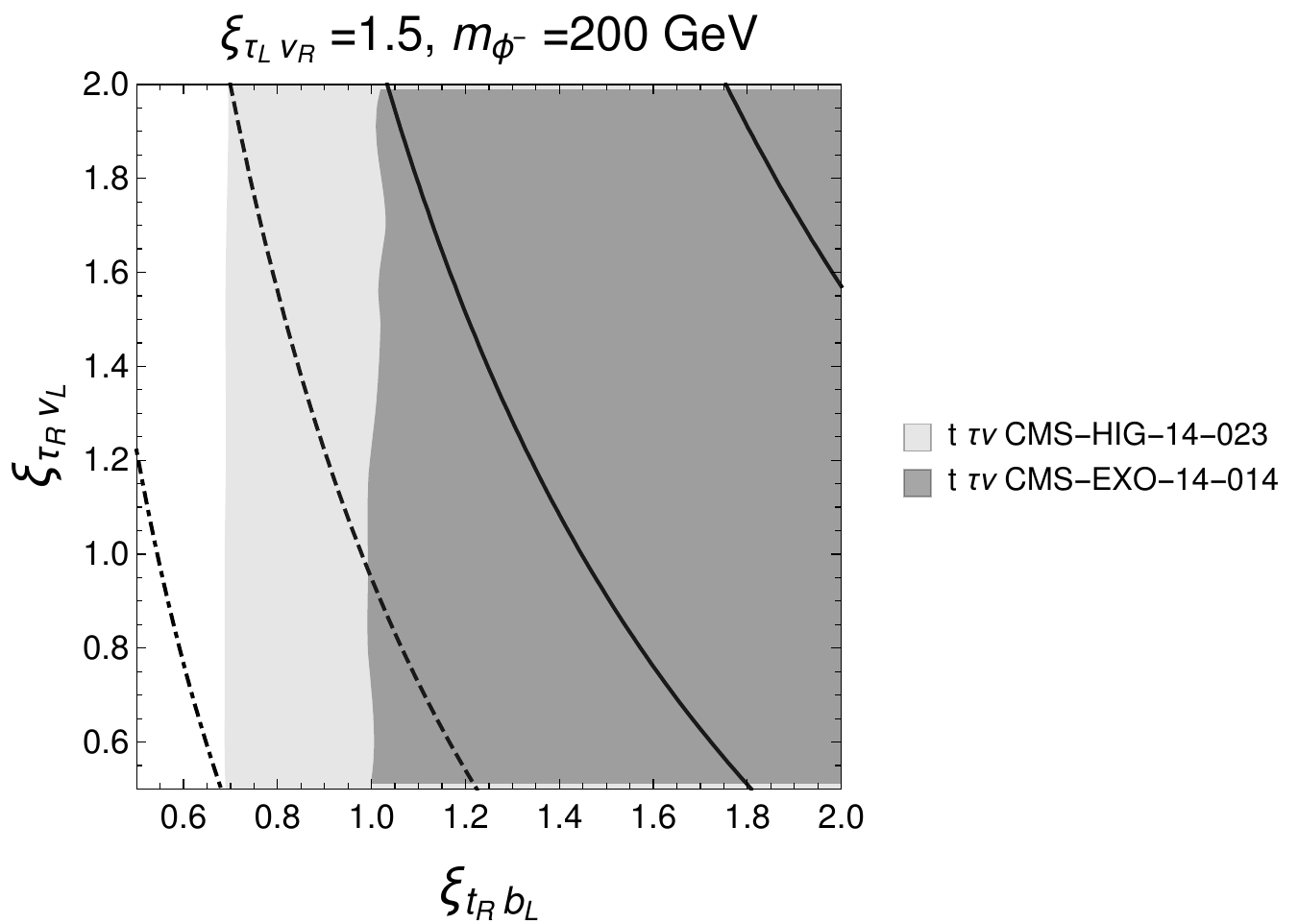} \nn 
	\end{eqnarray}
	\caption{Regions in grey represent the 95\% exclusion bounds due to the indicated final states. The areas enclosed by solid/dashed/dot-dashed lines represent the $1/2/3\sigma$ confidence regions for the correlated measurements of $R_{D}$ and $R_{D^{*}}$, which assume the values of the neutral scalar couplings indicated by the red star.}
	\label{fig:b2}
\end{figure}

\subsubsection*{Benchmark 1: $m_{\phi^0} = 200$ GeV $m_{\phi^-} = 200$ GeV} % (fold)

In Fig.~\ref{fig:b2} we show the bounds that LHC searches impose once we assume a mass of 200 GeV for both the charged and neutrals scalar fields of the simplified model. As we can see in the top panels, the most stringent bounds on the couplings of $\phi^0$ are due to the single production of the particle in association with a top quark. The allowed leptonic final states result in comparable 95\% exclusion regions, which force the transverse coupling to $t$ and $c$ below unity. The double production of $\phi^0$ results in a weaker  constraint that is only marginally relevant once the $\tau$ decay channel is predominant. 
The bottom panels refer instead to the charged scalars $\phi^\pm$. Also in this case we find that the single production in concomitance with an (anti)top quark casts the strongest bound, limiting the value of the coupling to $t$ and $b$ quarks below about 0.7. In these plots we show also the confidence regions designated by the measurements of $\rdsb$, where we assume the values of the neutral scalar couplings indicated by a red star in the top panels. 

\subsubsection*{Benchmark 2: $m_{\phi^0} = 300$ GeV $m_{\phi^-} = 300$ GeV} % (fold)

\begin{figure}[H]
	\begin{eqnarray}
	&&\includegraphics[scale=0.5]{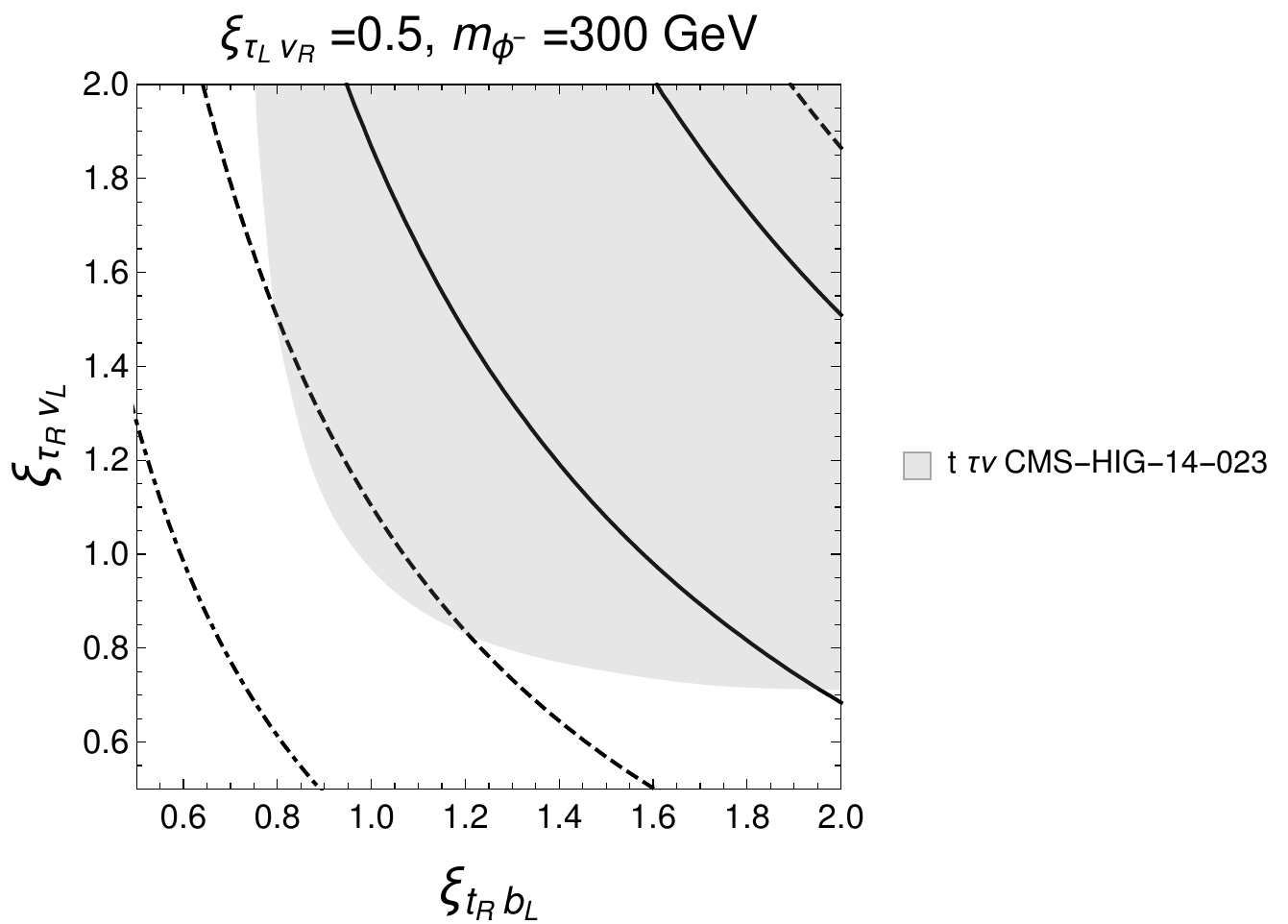}\hspace{2.1cm}
	\includegraphics[scale=0.5]{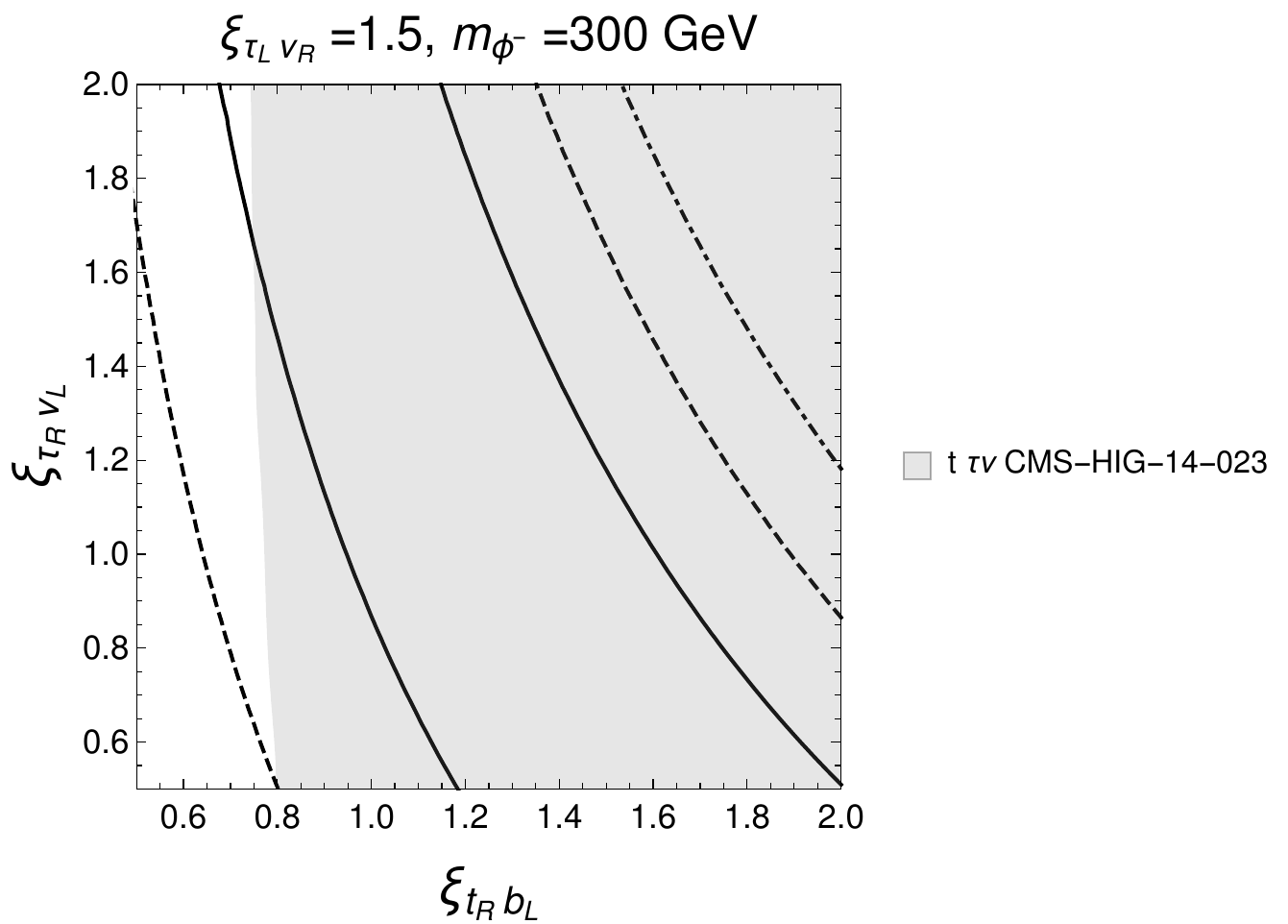} \nn 
	\end{eqnarray}
	\caption{Regions in grey represent the 95\% exclusion bounds due to the indicated final states. The areas enclosed by solid/dashed/dot-dashed lines are the $1/2/3\sigma$ confidence region for the correlated measurements of $R_{D}$ and $R_{D^{*}}$ assuming $\xi_{c_L t_R} = \xi_{\tau_R \tau_L} = 1.95$.}
	\label{fig:b3}
\end{figure}

Considering larger values for the scalar masses, the constraints in the top panels of Fig.~\ref{fig:b2} significantly relax and the simplified model in Eq.~\eqref{LagNP} is compatible with the measured values of $\rdsb$. As shown in Fig.~\ref{fig:b3}, while the properties of the charged scalar are still constrained by the dedicated searches, the current $B$ physics measurements can be explained for values of the couplings smaller than 2. For the computation of the confidence regions we here assumed $\xi_{c_L t_R} = \xi_{\tau_R \tau_L} = 1.95$. 

\subsubsection*{Benchmark 3: $m_{\phi^0} = 300$ GeV $m_{\phi^-} = 400$ GeV} % (fold)
 
\begin{figure}[H]
	\begin{eqnarray}
	&&\includegraphics[scale=0.5]{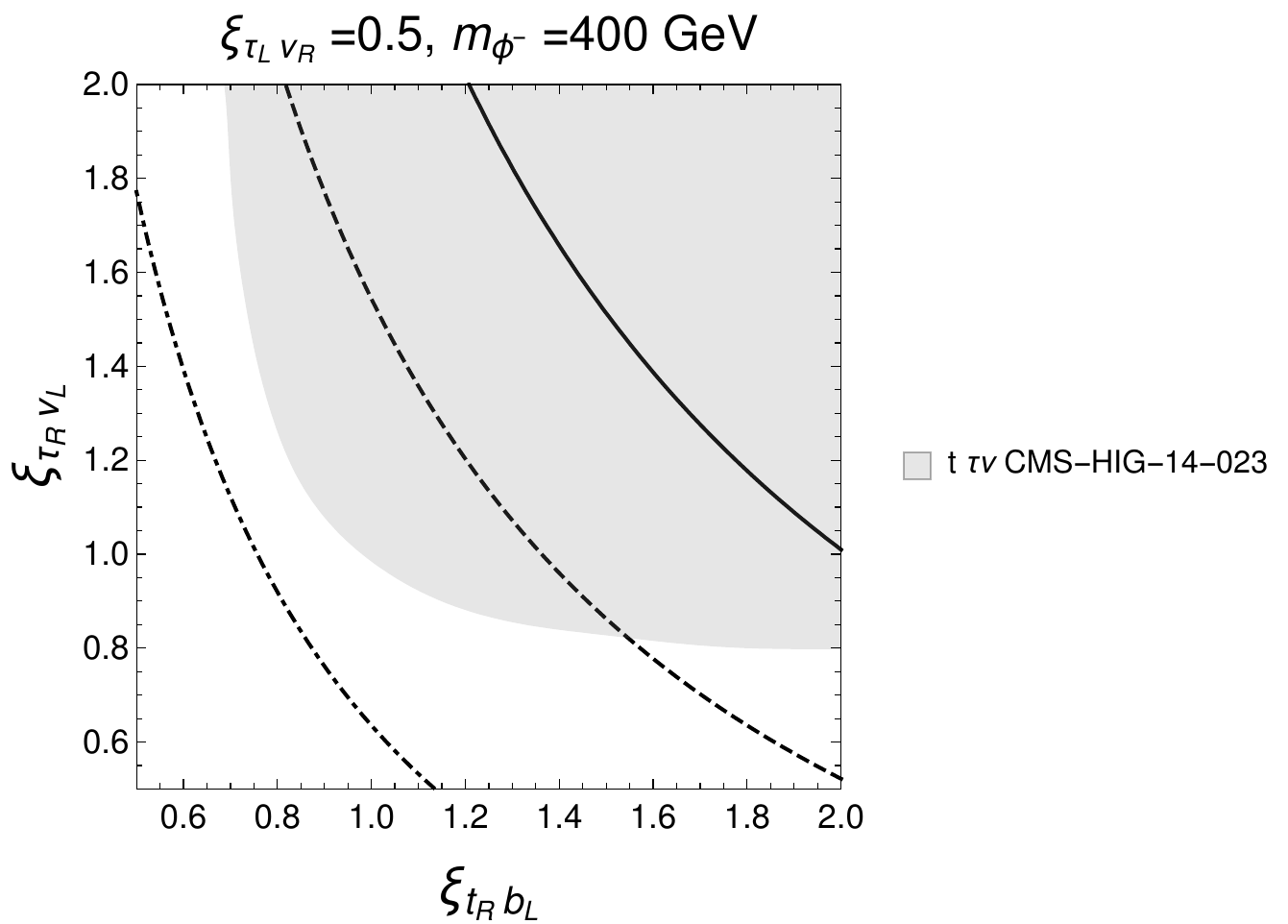}\hspace{2.1cm}
	\includegraphics[scale=0.5]{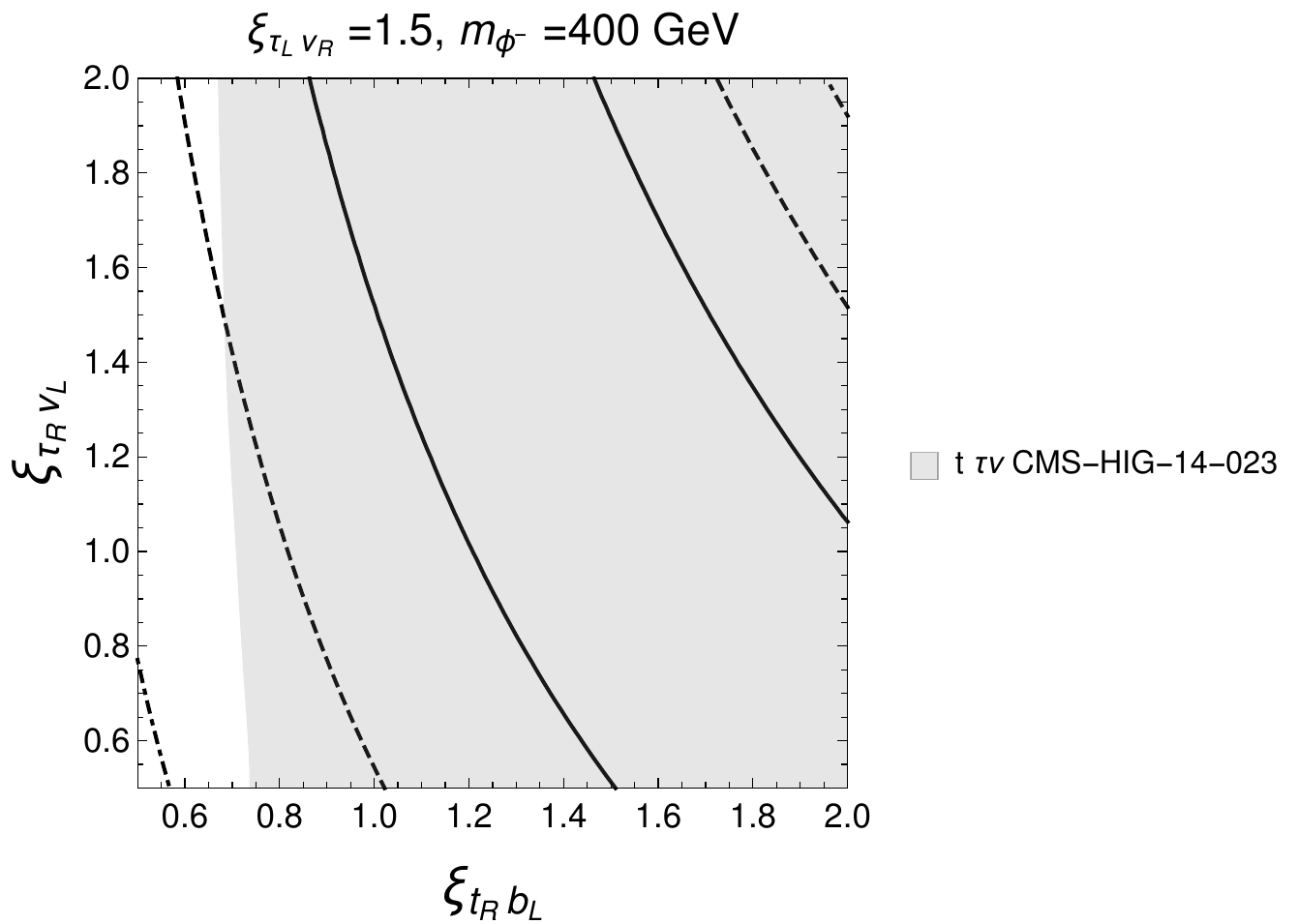} \nn 
	\end{eqnarray}
	\caption{Regions in grey represent the 95\% exclusion bounds due to the indicated final states. The areas enclosed by solid/dashed/dot-dashed lines are the $1/2/3\sigma$ confidence region for the correlated measurements of $R_{D}$ and $R_{D^{*}}$ assuming $\xi_{c_L t_R} = \xi_{\tau_R \tau_L} = 1.95$.}
	\label{fig:b4}
\end{figure}

As shown in Fig.~\ref{fig:b4}, larger values of the charged scalar mass lower the agreement of the model with the measurements of $\rdsb$. The latter in fact requires larger values of the couplings while the bounds of dedicated searches become progressively stricter~\cite{Khachatryan2015}.  

%-------------------------------------------------------------------------------
\section{A possible high-energy completion} % (fold)
\label{sec:A possible UV completion}
%-------------------------------------------------------------------------------

We investigate now a possible high-energy completion of our simplified model for flavour anomalies proposed in Eq.~\eqref{LagNP}. As a first attempt, we could identify the new scalar fields with the components of an $SU(2)$ doublet and immediately recover the invariance of the Lagrangian under the full SM gauge group. In this case, however, the SM quark mixing induced by the CKM matrix inevitably results in the dilution of the quark couplings to either the charged or the neutral scalar fields, diminishing the net contribution of the diagrams in Fig.~\ref{fig:S4}. This problem can be circumvented by assigning the new scalars to two \textit{different} $SU(2)$ doublets, effectively embedding the framework into a three-Higgs-doublets model (3HDM).
Proceeding along the lines of Ref.~\cite{Cordero:2017owj}, we therefore consider

\begin{eqnarray}
H_h = \left(\begin{array}{c}
	G^+ \\
	\frac{1}{\sqrt{2}}\left(v + h + i\,G^0 \right)
\end{array} \right)\,,\,\,\,
H_1 = \left(\begin{array}{c}
H_1^+ \\
H_1^0
\end{array} \right)\,,\,\,\,
H_2 = \left(\begin{array}{c}
H_2^+ \\
H_2^0
\end{array} \right)\,,
\label{doublets}
\end{eqnarray}

where $H_h$ plays the role of the SM Higgs doublet and is solely responsible for spontaneous symmetry breaking with a VEV of $v = 246$ GeV. The doublets $H_{1\,,2}$ instead provide the new scalar degrees of freedom which interact sizeably with the top quark, $\tau$ lepton and neutrinos.

In order to embed the Lagrangian of Eq.~\eqref{LagNP} in a 3HDM, we need to address the presence of two additional complex fields that can cause further conflicts with the LHC and low energy measurements, even though not directly involved in the $b\rightarrow c \tau \nu$ process. To estimate the mass scale of these states, we consider the potential
\begin{eqnarray}
V_{3HDM} &=& - \mu_h^2 H_h^{\dagger}\,H_h - \mu_1^2 H_1^{\dagger}\,H_1 - \mu_2^2 H_2^{\dagger}\,H_2 \nn \\
&+& \lambda_{11}\,(H_1^{\dagger}\,H_1)^2 + \lambda_{22}\,(H_2^{\dagger}\,H_2)^2 + \lambda_{hh}\,(H_h^{\dagger}\,H_h)^2\nn\\
&+& \lambda_{12}\,(H_1^{\dagger}\,H_1)(H_2^{\dagger}\,H_2) + \lambda_{2h}\,(H_2^{\dagger}\,H_2)(H_h^{\dagger}\,H_h) + \lambda_{h1}\,(H_h^{\dagger}\,H_h)(H_1^{\dagger}\,H_1)\nn\\
&+& \lambda_{12}'\,(H_1^{\dagger}\,H_2)(H_2^{\dagger}\,H_1) + \lambda_{2h}'\,(H_2^{\dagger}\,H_h)(H_h^{\dagger}\,H_2) + \lambda_{h1}'\,(H_h^{\dagger}\,H_1)(H_1^{\dagger}\,H_h)\, ,
\end{eqnarray} 
resulting in
\begin{eqnarray}
&& m^2_{H^0_1} = \left(-\mu_1^2 + \frac{1}{2}\left(\lambda_{h1} + \lambda_{h1}' \right)v^2 \right)\,, \; m^2_{H^{\pm}_1} = \left(-\mu_1^2 + \frac{1}{2}\left(\lambda_{h1}\right)v^2 \right),\nn\\
&& m^2_{H^0_2} = \left(-\mu_2^2 + \frac{1}{2}\left(\lambda_{2h} + \lambda_{2h}' \right)v^2 \right)\,, \; m^2_{H^{\pm}_2} = \left(-\mu_2^2 + \frac{1}{2}\left(\lambda_{2h}\right)v^2 \right)\,.
\end{eqnarray}
The mass-splitting between the neutral and charged elements of a doublet is  controlled by $\lambda'_{h1/2h}\,v^2/2$, therefore the mass degeneracy can be broken by a factor of about $v$.

The interactions of the new states follow form the Lagrangian

\begin{equation}
	- \mathcal{L}_{UV} \supset \bar{Q}_L \tilde{H}_1 \mathcal{Y}_1^u u_R + \bar{Q}_L \tilde{H}_2 \mathcal{Y}_2^u u_R + \bar{L}_L H_1 \mathcal{Y}_1^{\tau} \tau_R + \bar{L}_L H_2 \mathcal{Y}_2^{\tau} \tau_R + \bar{L}_L \tilde{H}_1 \mathcal{Y}_1^{\nu} \nu_R + \bar{L}_L \tilde{H}_2 \mathcal{Y}_2^{\nu} \nu_R + h.c. \, ,
	\label{SU2}
\end{equation}

where we identified the lightest components of each doublet with the scalar fields used in our simplified model: $\phi^0 \equiv H_1^0$, $\phi^- \equiv H_2^-$. The interactions that characterise the latter can then be recovered by  appropriately choosing Yukawa couplings of the form
\begin{eqnarray}
\mathcal{Y}_1^u =\left( \begin{matrix}
	0	& 0 & 0 \\ 
	0	& 0 & \xi_{\bar{c}_L t_R} \\ 
	0	& 0 & 0 
\end{matrix}\right)\,,
\quad
\mathcal{Y}_2^u = V^{ckm} \tilde{\mathcal{Y}}_2^u = V^{ckm}\left( \begin{matrix}
	0	& 0 & 0 \\ 
	0	& 0 & 0 \\ 
	0	& 0 & \xi_{\bar{t}_R b_L} 
\end{matrix}\right)\,,
\end{eqnarray}
written on the basis of the quark mass eigenstates. After having performed the CKM rotation, the scalars $H_1^0$ and $H_2^-$ couple then as
\begin{eqnarray}
-\mathcal{L} \supset H_1^0\,\bar{u}_L\,\mathcal{Y}_1^u u_R + H_2^- \bar{d}_L \tilde{\mathcal{Y}}_2^u u_R + h.c.\,,
\label{Alig}
\end{eqnarray}
matching the interactions in Eq.~\eqref{LagNP}. The heavier components of the doublets interact instead according to 
\begin{eqnarray}
-\mathcal{L}_H = H_2^0\,\bar{u}_L\,V^{ckm}\tilde{\mathcal{Y}}_2^u u_R + H_1^- \bar{d}_L V^{ckm\,\dagger} \mathcal{Y}_1^u u_R + h.c. \, ,
\label{Rot}
\end{eqnarray}
presenting suppressed couplings with respect to those of $H_1^0$ and $H_2^-$.

Although the collider signatures of the heavy components are thus less prominent than those of their light counterparts\footnote{The presence of additional gauge interaction does not worsen the bounds found in the previous section, which are mainly driven by single production processes that gauge interactions do not enhance. }, the new couplings contained in Eq.~\eqref{Rot} potentially induce flavour changing neutral currents and source the $B\rightarrow X_s \gamma$ process. In the following we investigate the constraints that the corresponding searches yield, and use these results to bound the mass spectrum of the full model.

\subsection{$\mathbf{B\to X_s} \boldsymbol{\gamma}$}

\begin{figure}[t]
	\centering
	\includegraphics[width=0.35\linewidth]{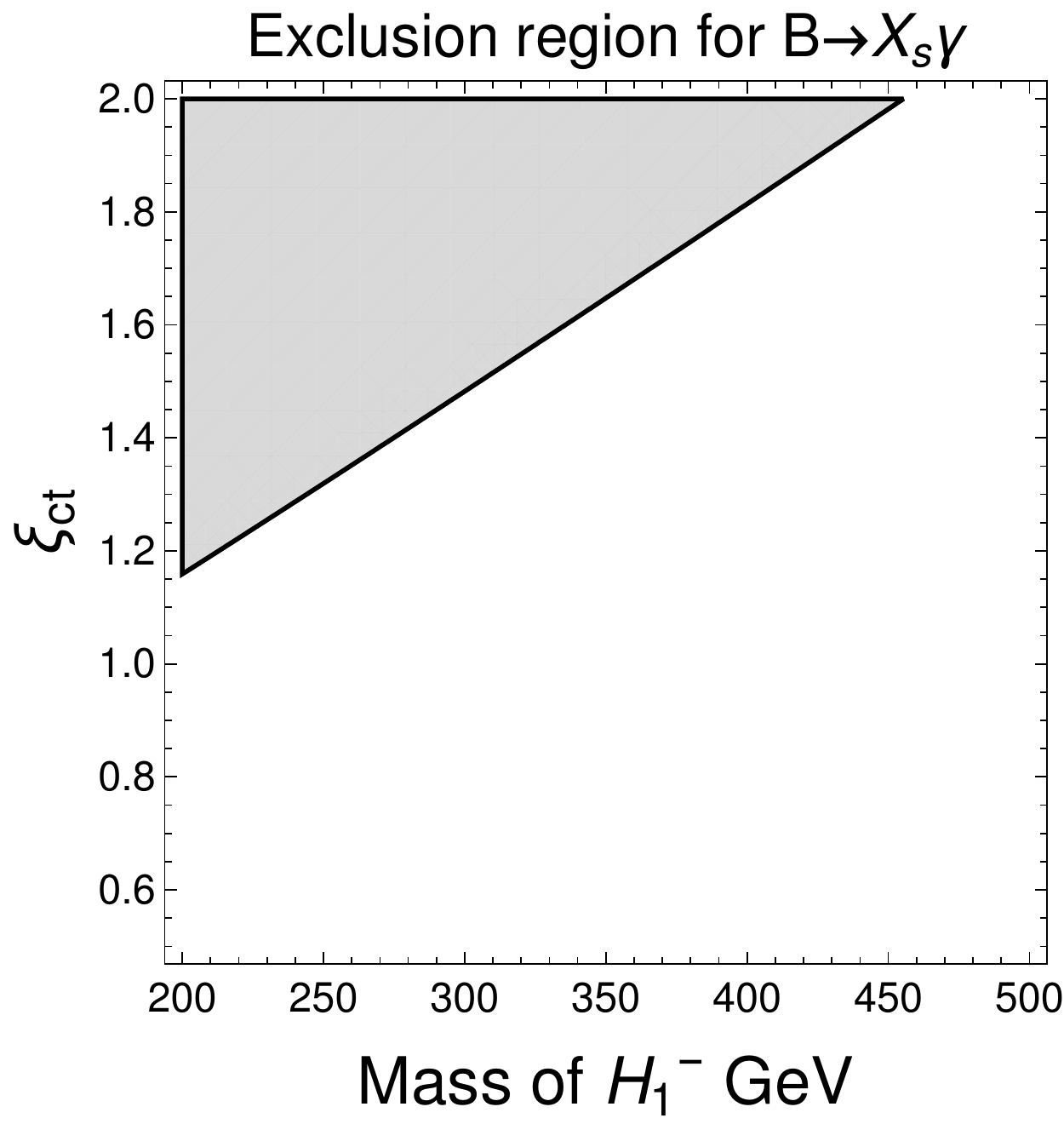}
	\caption{The grey area represents the exclusion bound imposed on new physics contributions to the magnetic dipole operators of gluon and photons at the one-loop level by observations of the process $B\to X_s \gamma$.}
	\label{fig:BsG}
\end{figure}

The charged current interactions in Eq.~\eqref{Rot} induce new couplings of the top quark to the down-type quarks, resulting in a non-zero contribution to the photon and gluon magnetic dipole operators\footnote{We neglect the  operators $P^{'}_7$ and $P^{'}_8$ with opposite chirality because  of the overall $m_s/m_b$ suppression factor.}  
\begin{eqnarray}
&&P_7=\frac{e}{16\pi^2}m_b\left(\bar{s} \sigma^{\mu \nu} P_R b\right) F_{\mu \nu} \,, \nn \\
&&P_8=\frac{g_3}{16\pi^2}m_b\left(\bar{s} \sigma^{\mu \nu}\,T^a P_R b\right) G^a_{\mu \nu}\,.
\end{eqnarray}
By matching Eq.~\eqref{Rot} with the effective Lagrangian~\cite{Bobeth:1999ww}
\begin{eqnarray}
\mathcal{L}_{eff} = \mathcal{L}^{5 flavors}_{\small{QCD \times QED}} + \frac{4 G_F}{\sqrt{2}}\,V_{ts}{}^*\,V_{tb} \left(C_7 P_7 + C_8 P_8\right),
\label{EffBsG}
\end{eqnarray}
and requiring that the new physics contributions fall in the range~\cite{Hu:2016gpe,Misiak:2015xwa}
\begin{eqnarray}
-0.0634 \leq C_7 + 0.242 C_8 \leq 0.0464\,,
\end{eqnarray}
we obtain the exclusion bound on the $\xi_{\bar{c}_L t_R}$ coupling and on the mass of $H_1^-$ shown in Fig.~\ref{fig:BsG}.

We remark that similar to $H^0_1$, the interactions of $H_1^\pm$ with the up-type quarks involve only the top quark. Any sizeable contribution to $B\to X_s \gamma$ due to the exchange of light quarks is therefore absent.  

\subsection{Meson mixing}
\begin{figure}[H]
	\centering
	\includegraphics[scale=0.8]{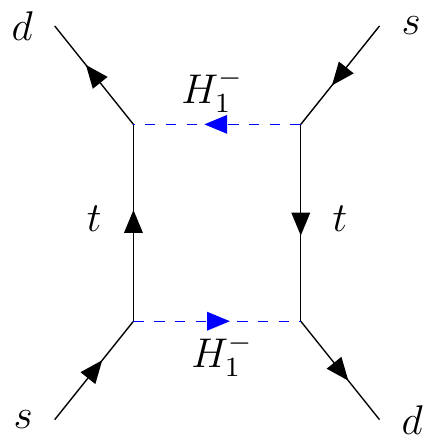} \hspace{.5cm}
	\includegraphics[scale=0.8]{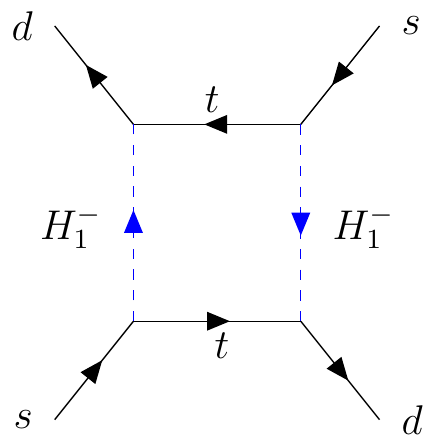}
	\caption{New diagrams contributing to the $K^0 - \bar{K}^0$ mixing.}
	\label{fig:Mes}
\end{figure}
Measurements of the mass splitting in the neutral $K$ system provide bounds on $\xi_{\bar{c}_L t_R}$ and the mass of $H_1^-$ which may compete with the constraint from $B\to X_s \gamma$. In regard of this, the interactions in Eq.~\eqref{SU2} contribute to $K^0 - \bar{K}^0$ via the two one-loop diagrams presented in Fig.~\ref{fig:Mes}. The value of the corresponding amplitude is to be matched with the effective Lagrangian
\begin{eqnarray}
- \mathcal{L}_{K} = z^K \left(\bar{d}_L \gamma_{\mu} s_L\right)\left(\bar{d}_L \gamma^{\mu} s_L\right)\, . 
\label{MesMixH}
\end{eqnarray}
Bounds on $z_1^K$ are then derived from measurements of the mass splitting $\Delta m_K/m_K = (7.01 \pm 0.01)\times 10^{-15}$ and of the involved CP-violating phase $\varepsilon_K = (2.23 \pm 0.01) \times 10^{-3}$, which respectively provide $|z^K| \le 8.8 \times 10^{-7}$ and $\Im{z_K}<3.3\times 10^{-9}$ \cite{Blum:2009sk}.  In our model we find 
\begin{eqnarray}
z^K \simeq - \frac{(V_{cs}^2 V^{*2}_{cd} \xi_{\bar{c}_L t_R}^4)}{16 \pi^2}\,D_{dd00}[m_t^2, m_t^2,m_{H_1^-}^2,m_{H_1^-}^2]\, ,
\end{eqnarray}
which is well in agreement with the considered limits on the considered parameter space.

\subsection{Lepton universality}
\begin{figure}[H]
	\centering
	\includegraphics[scale=0.7]{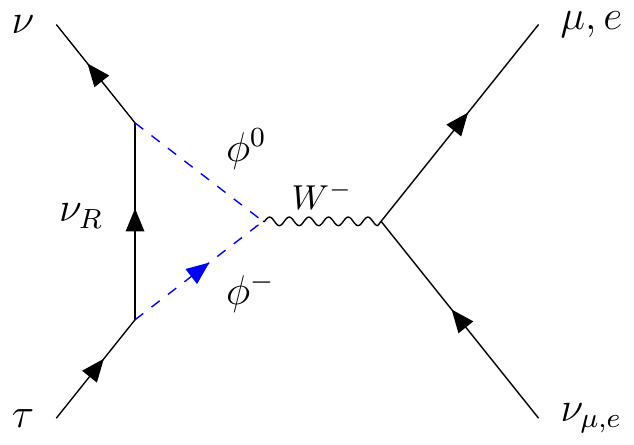}\hspace{.5 cm}
	\includegraphics[scale=0.7]{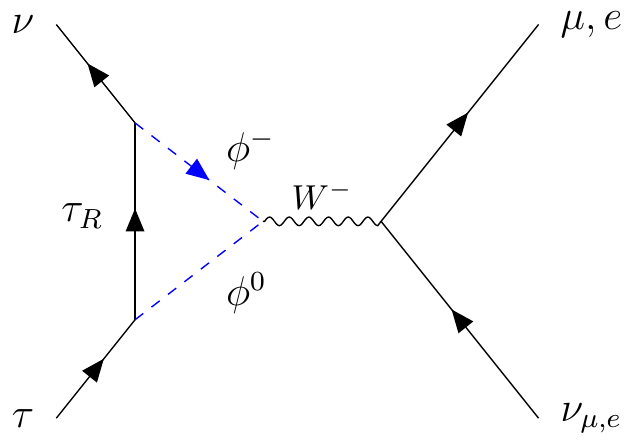}
	\caption{Diagrams leading to enhancements of the $\tau$ decay. $\phi^0$ and $\phi^-$ respectively represent the neutral and charged component of a scalar doublet.}
	\label{fig:G1}
\end{figure}
The coupling involving the third lepton generation in Eq.~\eqref{SU2} is subject to tight bounds from the decay of $\tau$ to lighter leptons mediated by gauge interactions in the full theory, shown in Fig.~\ref{fig:G1}. We then repeat for the model at hand the analysis in Ref.~\cite{Boucenna:2016qad}, assuming the data therein. The observables of reference are the two decay rates 
\begin{eqnarray}
\frac{\Gamma_{\tau \rightarrow e \nu \bar{\nu}}}{\Gamma_{\mu \rightarrow e \nu \bar{\nu}}} = \frac{|C^{\tau e}|^2}{|C^{\mu e}|^2} \times \frac{m_{\tau}^5 f(m_{e}^2/m_{\tau}^2)}{m_{\mu}^5 f(m_{e}^2/m_{\mu}^2)}\, ,\,\,\, \frac{\Gamma_{\tau \rightarrow \mu \nu \bar{\nu}}}{\Gamma_{\mu \rightarrow e \nu \bar{\nu}}} = \frac{|C^{\tau \mu}|^2}{|C^{\mu e}|^2} \times \frac{m_{\tau}^5 f(m_{\mu}^2/m_{\tau}^2)}{m_{\mu}^5 f(m_{e}^2/m_{\mu}^2)}\, ,
\label{tauF}
\end{eqnarray}
where $f(x)=1-8x+8x^3-x^4-12x^2 \ln(x)$ and a suitable normalization is used to cancel the dependence on $G_F$. 
The Wilson coefficient in Eq.~\eqref{tauF} are linked to the corresponding four leptons operator so that, in absence of new physics, the full SM effect is included in the Fermi constant. New physics contributions are then quantified by computing the induced finite shift of the weak coupling due to the extra scalar fields. At the leading order in the new Yukawa couplings, the divergences of the vertex are cancelled by the field strength renormalization of the $\tau$ and $\nu$ fields, leading to the finite contribution
\begin{eqnarray}
\delta g_2 = \sum_{i=1}^{2} \frac{\mathcal{Y}_i^{\tau}{}^2}{32\,\pi^2}\left(B_{bb1}(M_{H_i^-}) + B_{bb1}(M_{H_i^0}) + 4 C_{cc00}(M_{W};M_{H_i^0},M_{H_i^-}) \right) + \nn \\
\frac{\mathcal{Y}_i^{\nu}{}^2}{32\,\pi^2}\left(B_{bb1}(M_{\nu_R},M_{H_i^-}) + B_{bb1}(M_{\nu_R},M_{H_i^0}) + 4 C_{cc00}(M_{W};M_{\nu_R},M_{H_i^0},M_{H_i^-}) \right)\,,
\label{g2}
\end{eqnarray} 
and the Wilson coefficient
\begin{eqnarray}
C^{\ell\,\ell'} = \frac{4}{\sqrt{2}}G_F + \left(\frac{g_2^2}{2\,M_W^2}\right) \delta_{\ell,\tau} \times \delta g_2\, .
\end{eqnarray}
In the loop computation we neglect all external momenta and the mass of the $\tau$ lepton and neutrinos.
\begin{table}[h!]
	\centering
	\begin{tabularx}{1\textwidth}{X<\centering X<\centering X<\centering X<\centering}
		\toprule
		Observable & Experiment & SM prediction\\
		\hline
		$\Gamma_{\tau \rightarrow e \nu \bar{\nu}} / \Gamma_{\mu \rightarrow e \nu \bar{\nu}} $ & $1.350(4)\times10^6$  & $1.3456(5)\times10^6$ \\
		$ \Gamma_{\tau \rightarrow \mu \nu \bar{\nu}} / \Gamma_{\mu \rightarrow e \nu \bar{\nu}} $ & $1.320(4)\times10^6$  & $1.3087(5)\times10^6$ \\
	\bottomrule
	\end{tabularx}
	\caption{Measurements and SM predictions for the ratios $\Gamma_{\tau \rightarrow e \nu \bar{\nu}} / \Gamma_{\mu \rightarrow e \nu \bar{\nu}} $ and $ \Gamma_{\tau \rightarrow \mu \nu \bar{\nu}} / \Gamma_{\mu \rightarrow e \nu \bar{\nu}} $. The correlation coefficient is $0.45$ \cite{Boucenna:2016qad}.}
	\label{tab:Vicente} 
\end{table}

\begin{figure}[H]
	\centering
	\includegraphics[width=0.4\linewidth]{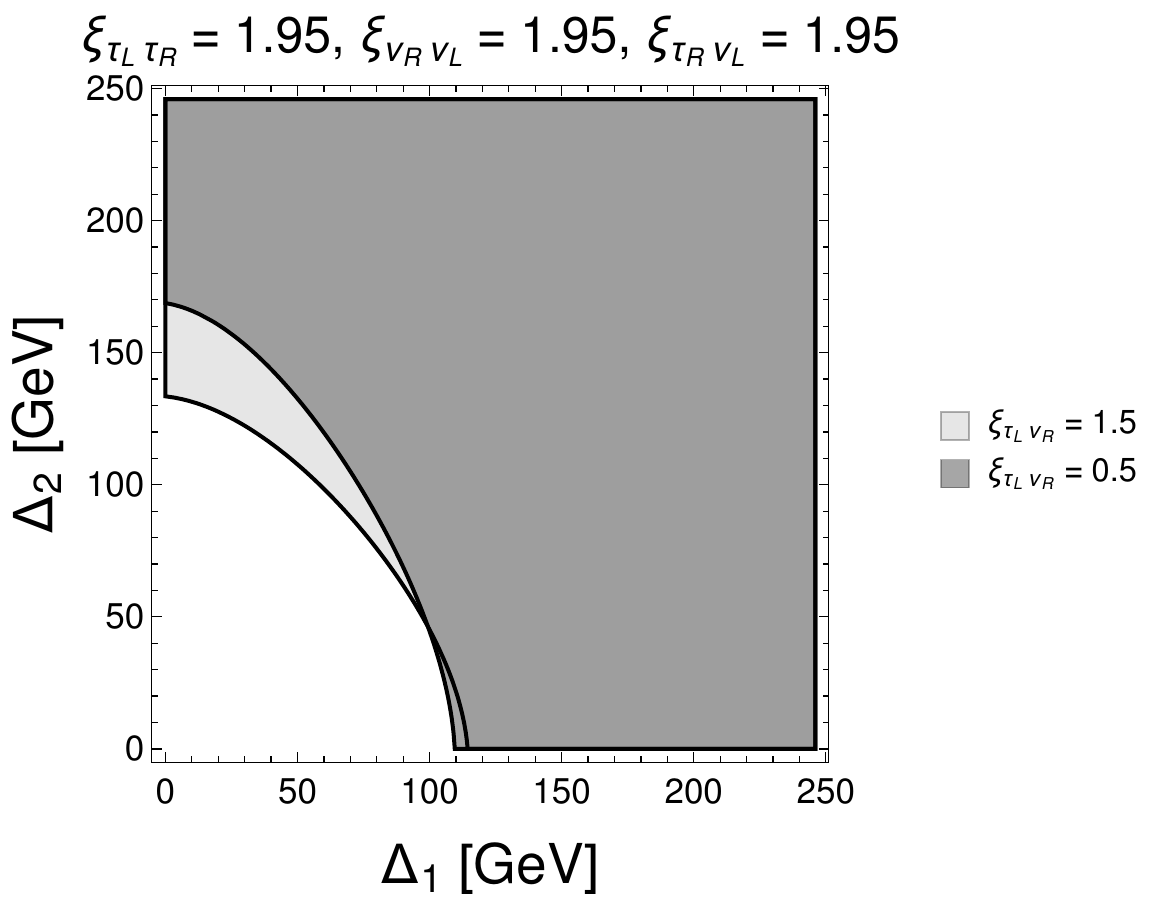}
	\includegraphics[width=0.4\linewidth]{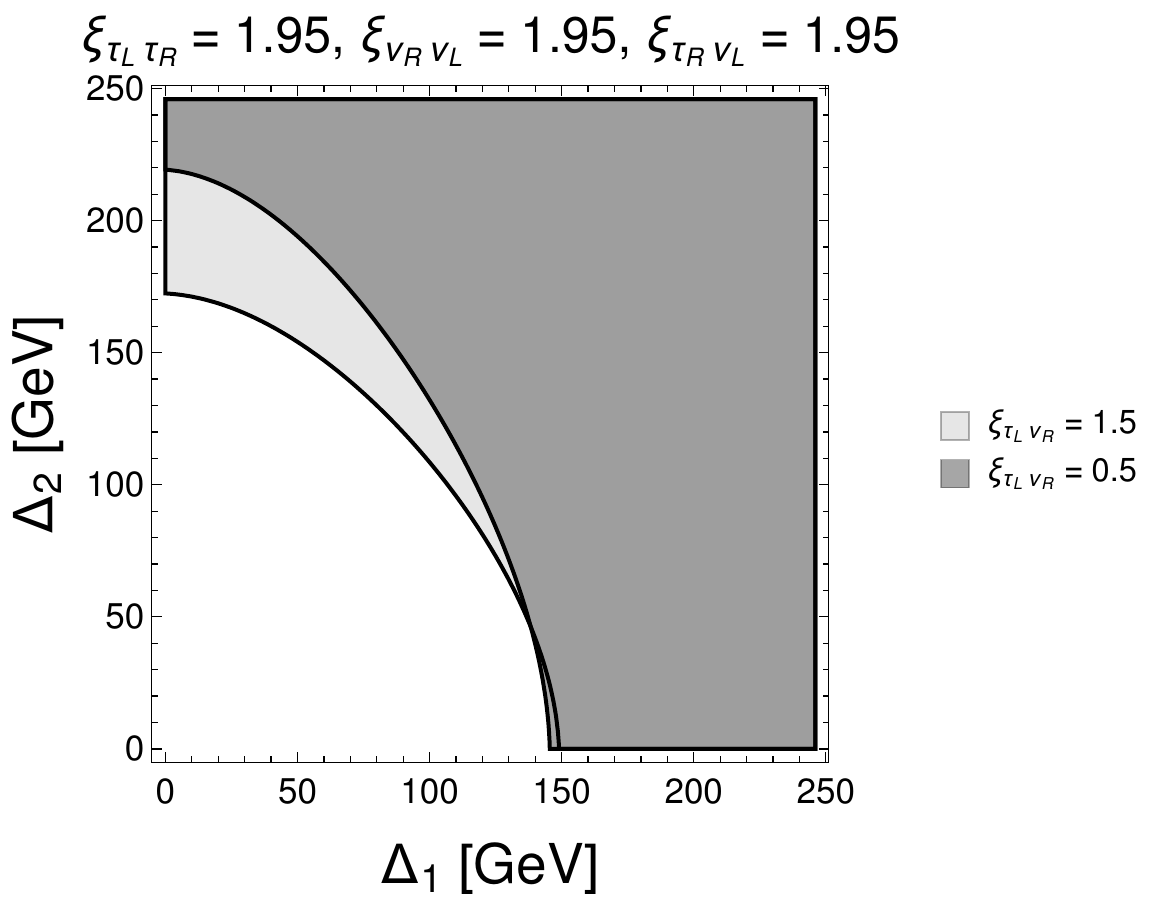}
	\caption{Exclusion regions ($\Delta\chi^2 > 3\sigma$) for $m_{\phi^0}=200$ GeV $m_{\phi^-}=200$ GeV and for $m_{\phi^0}=300$ GeV $m_{\phi^-}=300$ GeV. The values reported on the axes are the mass degeneracy factor of each doublet: $\Delta_1 = m_{H_1^-} - m_{H_1^0}$ and $\Delta_2 = m_{H_2^0} - m_{H_2^-}$.}
	\label{fig:tau}
\end{figure}

By comparing the result obtained with Eq.~\eqref{tauF} against the average of Table~\ref{tab:Vicente}, we find that our benchmark points are subject to the  exclusion regions presented in Fig.~\ref{fig:tau}, which constrain the mass splitting amongst the members of the two extra doublets. The more than $2\sigma$ discrepancy of the SM is always worsened by the interactions of non-degenerate doublets, therefore we refer to $3\sigma$ bounds in order to assess the exclusion regions of the model in Eq.~\eqref{SU2}.

\subsection{Electroweak precision tests}

We finally compute the contributions of the extra doublets to the $\rho$ parameter and compare it to the results of electroweak precision tests:

\begin{eqnarray}
\Delta_\rho = \frac{g_2^2}{64 \pi^2 M_W^2} \left[2\left( F(H_1^+, H_1^0) + F(H_2^+, H_2^0)\right) + 3\left( F(Z, h) - F(W, h) \right)\right]\,,
\end{eqnarray}

where~\cite{Grimus:2007if}

\begin{equation}
	F(x, y) = 
	\begin{dcases}
		\frac{m_x^2+m_y^2}{2} - 2\frac{m_x^2\,m_y^2}{m_x^2-m_y^2}\log\left(\frac{m_x}{m_y}\right), &\text{ if }x\neq y\,,\\
	0,	&\text{ if }x= y\,.
	\end{dcases}
\end{equation}

\begin{figure}[h]
  \centering
    \includegraphics[width=0.4\linewidth]{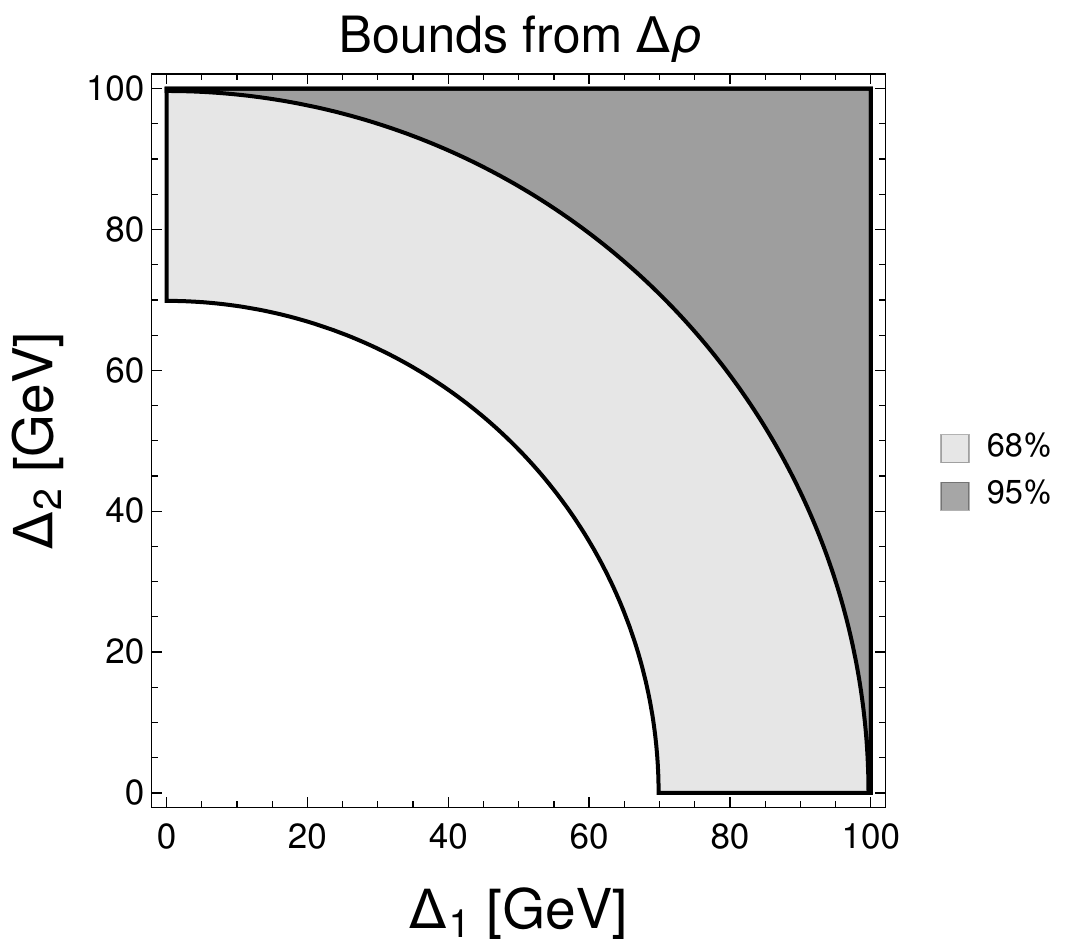}
  \caption{Exclusion regions due to precision measurements of the $\rho$ parameter.}
  \label{fig:Figures_DeltaRho}
\end{figure}

As shown in Fig.~\ref{fig:Figures_DeltaRho}, these measurements bound the mass splitting of the new doublets to $\Delta_{1,2}\lesssim 70$ GeV, regardless of the mass of the lightest state. 

%-------------------------------------------------------------------------------
\section{Conclusions} % (fold)
\label{sec:Conclusions}
%-------------------------------------------------------------------------------

If confirmed, the present B-physics anomalies would imply the existence of new physics beyond the Standard Model. Current provisional explanations seem to require exotic new physics, such as leptoquarks
characterised by very large couplings close to the perturbative limit. 

As an alternative, we propose a simple scalar extension of the SM that explains the observed deviations from lepton universality via a loop contribution that mimics the effect of vector degrees of freedom. The simplified scalar model we consider induces deviations only in vector Wilson coefficients, avoiding in this way the phenomenological bounds that affect traditional scalar extensions. We have shown that our proposal is phenomenologically consistent with all relevant collider as well as precision physics bounds. Our solution requires new scalars with couplings of order unity, thereby avoiding new non-perturbative physics close to the electroweak scale.

Considering the resulting loop contributions on top of the Standard Model prediction reduces the tension with the $B$ physics measurements below the $2\sigma$ level. We speculate that, once these loop effects are considered on top of a small tree level contribution, agreement with the anomalous signal could reach well into the $1\sigma$ range.
	
We also investigated a possible high-energy completion of the proposed simplified scalar model, identified in a three-Higgs doublet framework. In this case, measurements of lepton universality and of the $\rho$ parameters strongly constrain the mass splitting between the elements of the new scalar doublets. Given that explanation of the $\rdsb$ measurements require scalars slightly above the electroweak scale, we expect that future collider searches will  exhaustively probe the proposed scalar extension.

% section Conclusions (end)

%===============================================================================
% BACK PAGES
%===============================================================================

%-------------------------------------------------------------------------------
\section*{Acknowledgement} % (fold)

It is a pleasure to acknowledge  A. Crivellin, A. Djouadi and J. Martin Camalich for useful discussions. We also thank S. Di Chiara for participating in the early stage of this project. This work is supported by the Estonian Research Council grants MOBTT5, IUT23-6, PUT799, PUT808, the ERDF Centre of Excellence project TK133 and the European Research Council grant NEO-NAT.    

%-------------------------------------------------------------------------------

% section section name (end)

\bibliographystyle{hunsrt}
\bibliography{bib}

\end{document}